\documentclass[twocolumn,showpacs,superscriptaddress,preprintnumbers,amsmath,amssymb,prb,aps]{revtex4-1}
\usepackage{graphicx}
\usepackage{dcolumn}
\usepackage{bm}
\usepackage{multirow}
\begin{document}

\title{Spin-flip Raman scattering of the neutral and charged excitons confined in a CdTe/(Cd,Mg)Te quantum well}

\author{J. Debus}
\email[Corresponding author: ]{joerg.debus@tu-dortmund.de}
\affiliation{Experimentelle Physik 2, Technische Universit\"at Dortmund, 44227 Dortmund, Germany}

\author{D. Dunker}
\affiliation{Experimentelle Physik 2, Technische Universit\"at Dortmund, 44227 Dortmund, Germany}

\author{V. F. Sapega}
\affiliation{Ioffe Physical-Technical Institute, Russian Academy of Sciences, 194021 St. Petersburg, Russia}

\author{D. R. Yakovlev}
\affiliation{Experimentelle Physik 2, Technische Universit\"at Dortmund, 44227 Dortmund, Germany} \affiliation{Ioffe Physical-Technical Institute, Russian Academy of Sciences, 194021 St. Petersburg, Russia}

\author{G. Karczewski}
\author{T. Wojtowicz}
\author{J. Kossut}
\affiliation{Institute of Physics, Polish Academy of Sciences, 02668 Warsaw, Poland}

\author{M. Bayer}
\affiliation{Experimentelle Physik 2, Technische Universit\"at Dortmund, 44227 Dortmund, Germany}

\begin{abstract}
Spin-flip Raman scattering of electrons and heavy-holes is studied for resonant excitation of neutral and charged excitons in a CdTe/Cd$_{0.63}$Mg$_{0.37}$Te quantum well. The spin-flip scattering is characterized by its dependence on the incident and scattered light polarization as well as on the magnetic field strength and orientation. Model schemes of electric-dipole allowed spin-flip Raman processes in the exciton complexes are compared to the experimental observations, from which we find that lowering of the exciton symmetry, time of carrier spin relaxation, and mixing between electron states and, respectively, light- and heavy-hole states play an essential role in the scattering. At the exciton resonance, anisotropic exchange interaction induces heavy-hole spin-flip scattering, while acoustic phonon interaction is mainly responsible for the electron spin-flip. In resonance with the positively and negatively charged excitons, anisotropic electron-hole exchange as well as mixed electron states allow spin-flip scattering. Variations in the resonant excitation energy and lattice temperature demonstrate that localization of resident electrons and holes controls the Raman process probability and is also responsible for symmetry reduction. We show that the intensity of the electron spin-flip scattering is strongly affected by the lifetime of the exciton complex and in tilted magnetic fields by the angular dependence of the anisotropic electron-hole exchange interaction.
\end{abstract}


\maketitle

\section{Introduction}
The dynamics of carrier spins in low-dimensional semiconductor structures attract remarkable interest due to the possibilities of spin storage and transfer and related information processing\cite{AwschalomSpintronics,SemiconductorQu,Gabibook}. For spin electronic and quantum information applications, understanding of the fundamental interactions between confined carrier spins is essential, since these interactions may limit information handling due to spin relaxation\cite{spinbook}. An essential interaction between two confined carrier spins is the exchange interaction\cite{Kavokin,Kavokinb,Gorkov} arising from the carrier-carrier Coulomb interaction. The exchange interaction can be divided into isotropic and anisotropic contributions. While the isotropic exchange conserves the total spin of both carriers involved, the anisotropic one leads to spin relaxation. In that context, relaxation due to electron-hole exchange and due to exchange interaction between identically charged carriers (electrons or holes) need to be considered\cite{Bonesteel,Weng}. So far, exchange-related spin relaxation has been studied mostly for donor-bound electrons\cite{Harmon,Dzhioev,Kavokinc}. The effect of anisotropic exchange interactions between an exciton and a resident electron or hole localized by potential fluctuations in a quantum well (QW) has not been addressed in detail yet.

The resonant spin-flip Raman scattering (SFRS) technique offers the possibility to study carrier spins in semiconductor nanostructures\cite{Sapegab,Sirenkob,Koudinov}. By means of SFRS the $g$ factors of carriers can be measured, thus supporting the identification of the type of carriers involved in the scattering. In CdTe QWs the SFRS technique was used to measure the electron and exciton $g$ factors\cite{Sirenko,Kiselev,Shen}. Their dependences on the QW confinement and splitting between light-hole and heavy-hole states were determined. While the exciton-SFRS is arranged via simultaneous spin-flips of the constituents electron and hole, single carrier spin-flip of the electron or hole violates the conservation of the angular momentum in high-symmetry low-dimensional structures (e.g. with $D_{2d}$ symmetry). Possible mechanisms due to which the resonant Raman spin-flip of the electron or hole can nevertheless take place have not been discussed yet~\cite{Sirenko,Puls,Shen}.

In this paper, we report on the experimental study of SFRS in a CdTe/Cd$_{0.63}$Mg$_{0.37}$Te quantum well. The Raman scattering processes of the electron and heavy-hole spins are compared for resonant excitation of the neutral as well as positively and negatively charged excitons. We demonstrate that the spin-flip scattering of a single electron or hole in a neutral exciton becomes allowed when the exciton symmetry is reduced. As a result of the lifting of the angular momentum conservation, electrons and/or holes can mutually interact via anisotropic exchange. Also, a magnetic field tilted with respect to the QW growth axis can provide an electron- or a hole-SFRS process. The scattering via an acoustic phonon or isotropic exchange supports the symmetry-breaking in order to fulfill the required energy conservation. The strength and angular dependence of the electron-hole and hole-hole anisotropic exchange interactions and the underlying role of electron and hole localization in the QW for the SFRS are discussed.

The paper is organized as follows: in Sec.~\ref{expdetails} the studied sample and the experimental setup are specified. In Sec.~\ref{secpl}, the photoluminescence of neutral and charged excitons in dependence on the magnetic field is shown. In Sec.~\ref{secsfrs} the properties of the electron- and heavy-hole-SFRS in Faraday geometry are described. In Sec.~\ref{secangle} it is illustrated how the shift and intensity of the SFRS resonances vary with the angle between the magnetic field direction and QW growth axis. In the following Sec.~\ref{sectemp}, the impacts of lattice temperature and power density of above-barrier illumination on the electron- and hole-SFRS intensities are shown. Sec.~\ref{discussion} is concerned with model SFRS mechanisms for different geometries as well as basic information on carrier-carrier exchange interactions. In Sec.~\ref{discusschap}, the mechanisms of the experimentally determined electron and heavy-hole spin-flips are discussed for resonance excitation of the exciton and trions.

\section{Experimental details} \label{expdetails}
The studied CdTe/Cd$_{0.63}$Mg$_{0.37}$Te quantum well structure (\#090505AC), grown by molecular-beam epitaxy on a (100)-oriented GaAs substrate, consists of a single 20-nm-thick CdTe quantum well embedded between 120-nm-thick and 150-nm-thick Cd$_{0.63}$Mg$_{0.37}$Te barriers. Although the sample is nominally undoped, the QW contains resident carriers, in the majority holes, due to impurities in the barriers~\cite{Zhukov,Bartsch}. The concentration of the resident holes is about 10$^{10}$~cm$^{-2}$ as evaluated from magnetoreflectivity spectra of charged excitons~\cite{Astakhov}. It exceeds the resident electron concentration by at least one order of magnitude. Both carrier types can still coexist being spatially separated in the QW plane in localization islands. The concentration and even the type of the majority resident carriers can be tuned by additional illumination with photon energies exceeding the band gap of the Cd$_{0.63}$Mg$_{0.37}$Te layers of 2.26~eV.~\cite{Zhukov, Bartsch}

\begin{figure}[t]
\centering
\includegraphics[width=8.5cm]{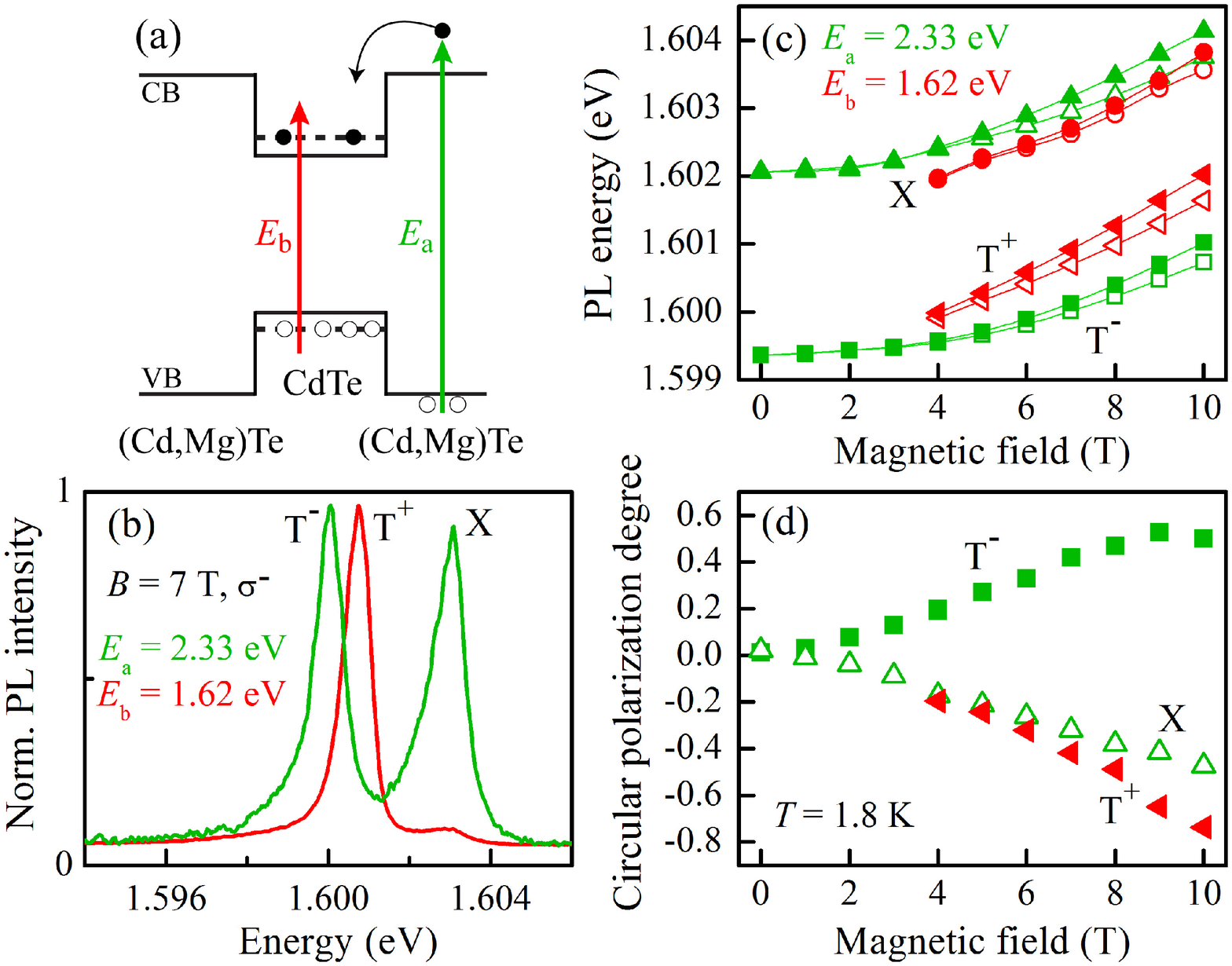}
\caption{\label{fig1} (Color online) (a) Band structure of the studied CdTe/(Cd,Mg)Te QW. Electrons (closed circles) and holes (open circles) are photogenerated by above- or below-barrier excitation with photon energies of $E_{\text{a}} = 2.33$~eV and $E_{\text{b}} = 1.62$~eV, respectively. (b) $\sigma^-$ circular-polarized PL spectra for linearly polarized above- and below-barrier excitation at $B = 7$~T in the Faraday geometry. $T=1.8$~K. Excitation densities are $P_{\text{a}} = 0.05$~W/cm$^2$ and $P_{\text{b}} = 0.8$~W/cm$^2$. (c) Magnetic field dependence of the PL line energies for both excitation energies. In high magnetic fields Zeeman splitting between $\sigma^+$ (closed symbols) and $\sigma^-$ (open symbols) circular-polarized components is observed for each line. (d) Circular polarization degree $\rho_{\text{c}}$ of the emission of neutral excitons and positively and negatively charged excitons {\sl vs} magnetic field.}
\end{figure}

Photogenerated electrons are collected from the Cd$_{0.63}$Mg$_{0.37}$Te layers into the CdTe QW, as illustrated in Fig.~\ref{fig1}(a). The holes remain in the barriers, being captured by surface states or trapping centers~\cite{Astakhov}. Depending on the illumination intensity two-dimensional electron gas densities up to a few 10$^{10}$~cm$^{-2}$ can be achieved. For below-barrier excitation electron-hole pairs are generated only in the CdTe QW and thus the density of the resident carriers is not changed. Such optical tuning of the resident carrier concentration is suitable for a comparative study of negatively and positively charged excitons in the same structure, avoiding technological variations if different samples would be examined. In the following, we will specify p-type and n-type regimes in the QW for the cases without and with above-barrier illumination, respectively.

The SFRS experiments were performed in high magnetic fields up to 10~T and at low temperatures ranging from 1.3 to 9~K. For resonant excitation of the neutral and charged exciton states a tunable continuous-wave (CW) Ti:Sapphire laser was used with a power density on the sample of about 0.5~W/cm$^2$. The additional above-barrier illumination was provided by a CW laser with a photon energy of 2.33~eV using power densities $P_\text{a}$ between 10$^{-4}$ and 15~W/cm$^2$. For below-barrier excitation a semiconductor laser with a photon energy of 1.62~eV was used. The laser beams for resonant excitation and illumination were directed to a joint position on the sample. The diameters of the laser spots were typically 0.5~mm. The circular or linear polarization of the exciting photons was selected by a Glan-Thompson prism combined with a quarter-wave or half-wave retardation plate. Corresponding polarization optics were used for analysis of the polarization of the Raman signal emitted from the sample.

The scattered light was dispersed by a triple-spectrometer (Princeton Instruments TriVista 555) consisting of three single 0.5~m spectrometers that can be operated in different modes. In the additive mode all spectrometer stages contribute to positive light dispersion with a total focal length of 1.5~m, thus providing a spectral resolution of about 0.03~meV for the slit width of 10~$\mu$m. The signal was detected by a photomultiplier. A charge-coupled-device (CCD) camera was used with the spectrometer operated in the subtractive mode. Here, the first two stages act as tunable bandpass filter providing high stray-light rejection.

Raman spectra were measured in backscattering geometry: incident and scattered light, propagating in opposite directions, were directed along the structure growth axis ($z$-axis). The magnetic field $\mathbf{B}$ was applied either along (Faraday geometry, $\mathbf{B} \parallel \mathbf{z}$) or perpendicular (Voigt geometry, $\mathbf{B} \perp \mathbf{z}$) to the growth axis. The SFRS signal was analyzed with respect to its circular polarization as indicated by the notation $z(\sigma^\eta, \sigma^\gamma)\bar{z}$ or in short $(\sigma^\eta, \sigma^\gamma)$.~\cite{Damen} Here, $z$ and $\bar{z}$ designate the optical $z$-axis and $\sigma^\eta$ and $\sigma^\gamma$ the circular polarizations of the incident and scattered light, respectively. The sign of $\eta$ and $\gamma$ is determined by the sign of the photon angular momentum projection on the optical axis. For angular dependent SFRS measurements, the angle $\theta$ between the magnetic field direction and $z$-axis was varied from 0$^\circ$ to 90$^\circ$. Since in tilted geometry the optical selection rules are lifted, the polarization was changed from circularly polarized to cross-linearly polarized. A crossed polarization configuration allowed us to strongly reduce the laser stray-light.

\section{Experimental results}
\subsection{Photoluminescence in magnetic field} \label{secpl}
The photoluminescence (PL) of neutral and charged excitons in the CdTe/Cd$_{0.63}$Mg$_{0.37}$Te QW is shown in Fig.~\ref{fig1}(b) for a magnetic field of $B=7$~T and $T=1.8$~K. Let us firstly consider below-barrier excitation at a photon energy of $E_{\text{b}} = 1.62$~eV. The weak PL line at 1.6031~eV stems from the recombination of heavy-hole (hh) excitons. The more intense line, shifted to lower energy relative to the exciton (X) one, is attributed to positive trion (T$^+$) recombination, see Ref.~\onlinecite{Bartsch} for details. When the sample is excited by laser light with $E_{\text{a}} = 2.33$~eV the type of resident carriers is tuned from holes to electrons~\cite{Zhukov,Bartsch}. Consequently, the negative trion (T$^-$) emission line emerges. It has a larger low-energy shift relative to the exciton emission as compared to the T$^+$ line. The energy differences between exciton and the trions correspond to the binding energies of T$^+$ and T$^-$.

In Figure~\ref{fig1}(c) the energies of the circular-polarized exciton and trion emission lines in external magnetic fields are shown for both below- and above-barrier excitations. They are contributed by the diamagnetic shift of the exciton complexes and the Zeeman splitting. The lines shift to higher energies with increasing magnetic field and experience an increasing splitting between the $\sigma^-$ (open symbols) and $\sigma^+$ (closed symbols) polarized components~\cite{Bartsch}.

In Figure~\ref{fig1}(d) the magnetic field dependences of the circular polarization degree of the exciton and trion emission are depicted. The polarization degree is calculated by $\rho_{\text{c}} = (I^{+} - I^{-})/(I^{+} + I^{-})$ from the PL intensities $I^{\pm}$ for $\sigma^+$ and $\sigma^-$ polarized light. It was found that in the studied QW both electron and heavy-hole $g$ factors have negative sign~\cite{Bartsch}. Therefore, the emissions from the singlet states of the T$^+$ and T$^-$ trions have opposite circular polarizations. The T$^+$ trion PL is dominantly $\sigma^-$ polarized, whereas the T$^-$ trion emission is more intense in $\sigma^+$ polarization. Thus, a negative $\rho_{\text{c}}$ is assigned to positive trions, while $\rho_{\text{c}} > 0$ is characteristic for negative trions.

\subsection{Spin-flip Raman scattering} \label{secsfrs}
\begin{figure}[t]
\centering
\includegraphics[width=6.7cm]{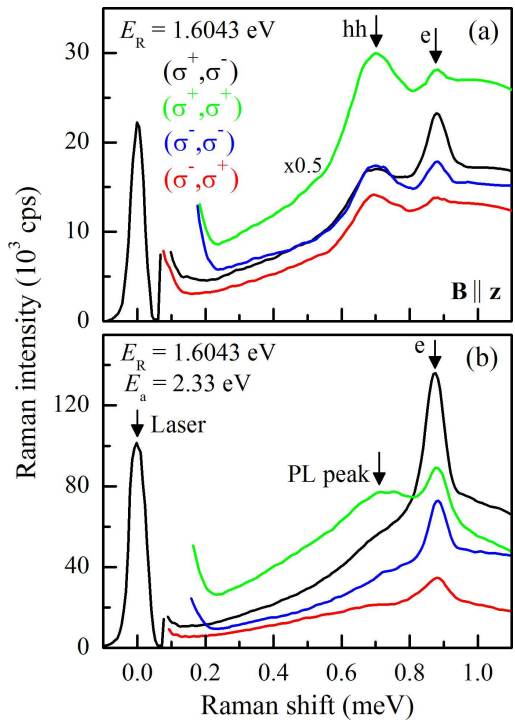}
\caption{\label{fig2} (Color online) SFRS Stokes-spectra in Faraday geometry for co- and counter-circular polarizations at $B = 9$~T and $T= 1.8$~K. (a) The exciton is resonantly excited at an energy of $E_\text{R}=1.6043$~eV. The laser line is at zero Raman shift. The strong intensity of the $(\sigma^{+},\sigma^{+})$ polarized SFRS spectrum is scaled down by a factor of 0.5 for better comparison. (b) The sample is additionally illuminated above barrier with excitation power $P_{\text{a}} = 0.04$~W/cm$^2$. Electron and heavy-hole spin-flip Raman lines as well as broad exciton PL are marked by arrows.}
\end{figure}

In Figure~\ref{fig2} SFRS Stokes-spectra for the co-circular $(\sigma^+, \sigma^+)$, $(\sigma^-, \sigma^-)$ and counter-circular $(\sigma^+, \sigma^-)$, $(\sigma^-, \sigma^+)$ polarization configurations are shown. The spectra were measured at $B=9$~T in Faraday geometry ($\mathbf{B} \parallel \mathbf{z}$) and $T=1.8$~K. The excitation energy $E_\text{R} = 1.6043$~eV is in resonance with high-energy side of the exciton PL. For this excitation two SFRS lines are observed on top of the PL background, see Fig.~\ref{fig2}(a). Their Raman shifts of 0.89 and 0.70~meV, reflecting Zeeman splittings of $|g| \mu_\text{B} B$ with the Bohr magneton $\mu_\text{B}$, correspond to the longitudinal $g$ factors $|g^{\parallel}_{\text{e}}|=1.71 \pm 0.01$ of the electron (e) and $|g^{\parallel}_{\text{hh}}|=1.34 \pm 0.02$ of the heavy-hole~\cite{Zhukov,Sirenko,Kiselev}.

\begin{figure}[t]
\centering
\includegraphics[width=6.7cm]{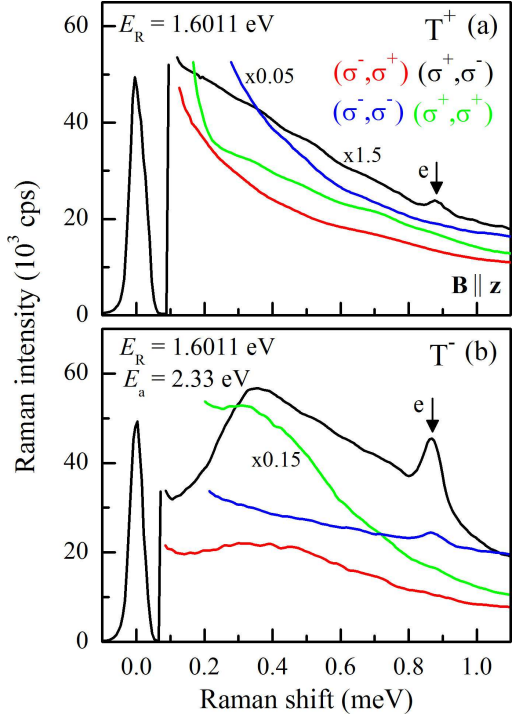}
\caption{\label{fig3} (Color online) SFRS Stokes-spectra in Faraday geometry measured at one of the trion resonances for co- and cross-circular polarizations at $B = 9$~T and $T = 1.8$~K. (a) Excitation of the positive trion yields an e-SFRS in the $(\sigma^{+},\sigma^{-})$ configuration. (b) Under above-barrier illumination at the T$^{-}$ resonance the e-SFRS is enhanced in the cross-polarized $(\sigma^{+},\sigma^{-})$ configuration. Under the illumination the trion PL is shifted to lower energies and its peak is seen at a Raman shift of about $0.4$~meV.}
\end{figure}

The e- and hh-SFRS lines are detected for both co- and cross-circular polarizations. While the e-SFRS is dominant in the $(\sigma^+, \sigma^-)$ configuration, the heavy-hole spin-flip is most intensive in $(\sigma^+, \sigma^+)$. The ratios of the SFRS intensities $I_\text{SF}$ for the polarization configurations $(\sigma^+, \sigma^-) : (\sigma^+, \sigma^+) : (\sigma^-,\sigma^-) : (\sigma^-, \sigma^+)$ at $E_\text{R} = 1.6043$~eV are given by
\begin{eqnarray} 
\text{X-resonance:} \label{intensratios}
\begin{cases}
\text{e-SFRS:} \quad \; \: \, 1:0.31:0.39:0.15 \\
\text{hh-SFRS:} \quad 0.27:1:0.33:0.28
\end{cases} \hspace{-0.35cm}.
\end{eqnarray}

The intensities of the e-SFRS line in the co-polarized configurations are about 1/3 of that in the $(\sigma^+, \sigma^-)$ configuration, while $I_\text{SF}(\sigma^-, \sigma^+)$ is only 1/7 of $I_\text{SF}(\sigma^+, \sigma^-)$ for the Stokes shifted lines. For the anti-Stokes processes (not shown here), where the scattered photon has a higher energy than the incident one, the e-SFRS line is most intensive in the $(\sigma^-, \sigma^+)$ configuration. The hh-SFRS has highest intensity for the $(\sigma^+, \sigma^+)$ configuration both for Stokes and anti-Stokes lines. In the other configurations $I_\text{SF}$ is reduced by a factor of 3 for the hh-SFRS line.

Above-barrier illumination changes the intensities of both SFRS lines, see Fig.~\ref{fig2}(b). The e-SFRS line is enhanced by about one order of magnitude, while the hh-SFRS line vanishes. This agrees well with our knowledge on the illumination effect, which provides resident electrons in the QW and, therefore, diminishes the concentration of the resident holes in favor of the electrons. The intensity ratios of the e-SFRS lines for the different polarization configurations are $1:0.36:0.42:0.20$. In comparison to the ratios in Eq.~(\ref{intensratios}), the illumination does not significantly change the relative intensities among the different configurations. Furthermore, the full width at half maximum (FWHM) of the e-SFRS line given by about 0.08~meV is not influenced by the illumination. This FWHM is mainly determined by the variation of the electron $g$ factor in the QW.

Moving the excitation laser to the trion resonances results in considerable modification of the SFRS spectra. In Figure~\ref{fig3}(a) the SFRS spectra for resonant excitation of the high-energy side of the T$^+$ PL are shown. The hh-SFRS line is absent in all polarizations and the e-SFRS is seen only in the $(\sigma^+,\sigma^-)$ configuration. Its intensity is approximately four times weaker than at the exciton resonance. In Figure~\ref{fig3}(b) the SFRS spectra with the illumination giving rise to resident electrons are shown. At the T$^-$ resonance the e-SFRS intensity is enhanced in cross-circular polarization $(\sigma^+, \sigma^-)$, and also a weak resonance is observed in $(\sigma^-, \sigma^-)$ configuration. On the anti-Stokes side (not shown here) the e-SFRS is detected in the opposite $(\sigma^-,\sigma^+)$ and $(\sigma^+,\sigma^+)$ configurations. Similar to the T$^{+}$ trion, the hh-SFRS is absent for the T$^{-}$ trion excitation. Worthwhile to remind here that in the studied sample without illumination a coexistence of the resident holes and electrons (with considerably smaller electron density compared to the hole one) has been established~\cite{Zhukov}. Therefore, the weak e-SFRS signal in Fig.~\ref{fig3}(a) may be related to T$^-$ trions.

\begin{figure}[t]
\centering
\includegraphics[width=6.8cm]{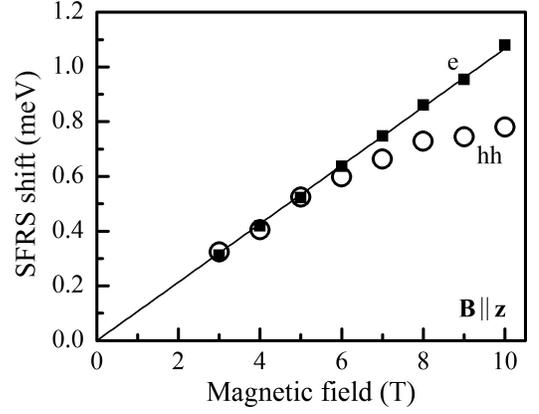}
\caption{\label{fig4} Magnetic field dependence of the spin-flip Raman shifts of electron and heavy-hole in Faraday geometry at $T = 1.8$~K. The resonant excitation energy is $E_\text{R} = 1.6043$~eV. The e-SFRS line (closed squares) shifts linearly with magnetic field, as illustrated by a linear fit (solid line). The hh-Raman shift (open circles) deviates from the linear behavior for $B > 5$~T.}
\end{figure}

The dependence of the electron and heavy-hole spin-flip Raman shifts on magnetic field in the Faraday geometry is plotted in Fig.~\ref{fig4}. Specific experimental conditions were chosen in order to avoid a spectral overlap of both SFRS lines for the excitation energy of 1.6043~eV. The hh-SFRS line is detected in the $(\sigma^+, \sigma^+)$ configuration and the e-SFRS is measured with illumination in the $(\sigma^+, \sigma^-)$ configuration, see Fig.~\ref{fig2}. The e-SFRS line (closed squares) shifts linearly with increasing magnetic field with a slope corresponding to the electron $g$ factor value of 1.71. The linear shift of the e-SFRS and its slope value are the same for resonant excitation of exciton and trion. In low magnetic fields ($B \leq 5$~T) the shift of the hh-SFRS line (open circles) within the experimental error is identical to that of the electron. In these fields Zeeman splittings and, correspondingly, longitudinal $g$ factors of electrons and heavy-holes are equal. For magnetic fields exceeding 5~T the hh-SFRS shift deviates from the linear dependence tending toward saturation. In such high magnetic fields the mixing of heavy-hole and light-hole (lh) states becomes essential~\cite{Bartsch}, leading to a reduction of the heavy-hole $g$ factor.

\begin{figure}[t]
\centering
\includegraphics[width=7.4cm]{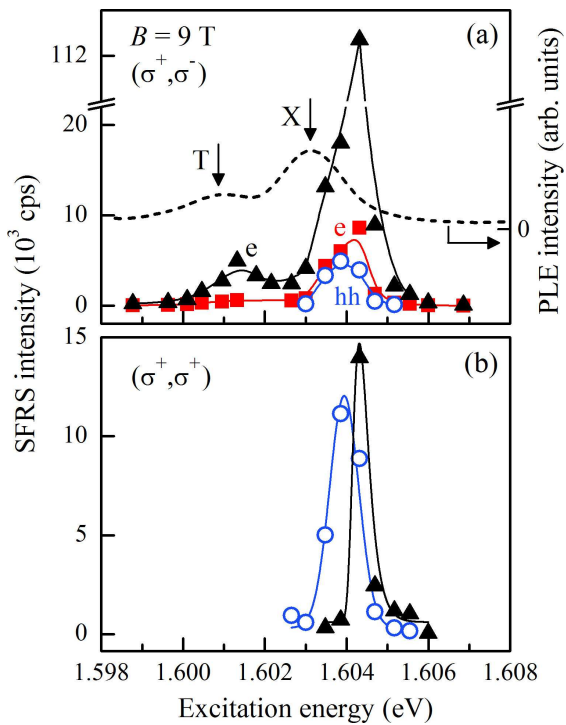}
\caption{\label{fig5} (Color online) SFRS resonance profiles of electron and heavy-hole in Faraday geometry at $B=9$~T and $T = 1.8$~K. The electron-SFRS line is measured with (black triangles) and without (red squares) above-barrier illumination. The heavy-hole SFRS (blue circles) is obtained for resonant excitation of the exciton. The solid lines are guides to the eye. A PL excitation spectrum detected at 1.5977~eV and at $B=9$~T is shown by the dashed line. The exciton and trion peaks are marked by vertical arrows. (b) SFRS resonance profiles for electron and heavy-hole measured in $(\sigma^+, \sigma^+)$ configuration. Experimental conditions and symbols are same as in panel (a).}
\end{figure}

The excitation spectrum of the electron- and heavy-hole-SFRS is illustrated in Fig.~\ref{fig5}(a). Here, changes in the intensity of the SFRS lines by varying the excitation energy across the exciton and trion resonances are shown for cross-circular polarization $(\sigma^+, \sigma^-)$. Profiles for the e-SFRS with (black triangles) and without (red squares) illumination as well as for the hh-SFRS (blue circles) without illumination are measured in the Faraday geometry at $B=9$~T. The hh-SFRS  has maximal intensity at $E_\text{R} = 1.6039$~eV and is absent across the entire trion resonance. The e-SFRS shows a peak at 1.6043~eV and possesses a second maximum at 1.6013~eV, which is about fifteen times smaller than the high-energy maximum (note the break of the intensity scale). The resonance profiles also indicate that the e-SFRS intensities for resonant excitation of the exciton or trion states are enhanced by illumination, but their energy positions do not change.

For comparison, a photoluminescence excitation (PLE) spectrum measured at $B=9$~T and $T=1.8$~K is shown by the dashed line in Fig.~\ref{fig5}(a). The PLE peaks of exciton and trion do not coincide with the SFRS resonance profile maxima. The SFRS intensities are maximal at the high-energy flanks of both PLE peaks. At these energies weakly localized states are probed~\cite{AKavokin}. Moreover, the shift of the e-SFRS profiles at the exciton resonance corresponds to the electron Zeeman splitting of about 1~meV at $B=9$~T. This demonstrates that the SFRS is more efficient for the outgoing scattering channel (exciton recombination), when the energy of the scattered photon coincides with the exciton transition.

The spectral dependences of the e-SFRS and hh-SFRS intensities in the co-polarized configuration $(\sigma^+, \sigma^+)$ are compared in Fig.~\ref{fig5}(b). Only in resonance with the exciton both SFRS lines are observed. The hh-SFRS shows a resonance profile spectrally similar to that in $(\sigma^+, \sigma^-)$, while the width of the e-SFRS resonance profile is narrower.

\subsection{Angular dependence of SFRS} \label{secangle}
\begin{figure}[t]
\centering
\includegraphics[width=8.4cm]{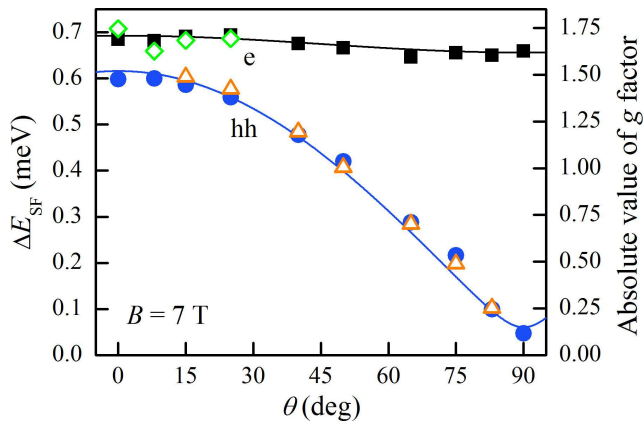}
\caption{\label{fig6} (Color online) Angular dependence of the spin-flip Raman shifts at $B=7$~T and $T=1.8$~K. The experimental error of the Raman shift does not exceed the symbol size. The e-SFRS (closed squares), measured at exciton resonance, shows only a slight anisotropy, while the Raman shift of the heavy-hole line (closed circles) strongly depends on tilting of the magnetic field direction. The solid curves represent fits based on Eq.~(\ref{eqgfactor}). The intensities of e- and hh-SFRS (open symbols), recorded at T$^+$ resonance, vary strongly with $\theta$. In particular, for $\theta > 25^\circ$ the e-SFRS line disappears. The angle-dependent e- and hh-SFRS shifts coincide for exciton and trion excitations.}
\end{figure}

\begin{figure}[t]
\centering
\includegraphics[width=6.2cm]{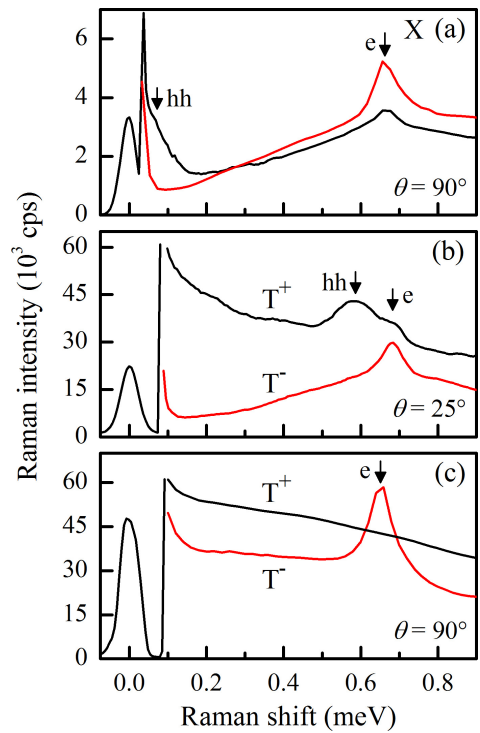}
\caption{\label{fig7} (Color online) SFRS spectra measured with (red line) and without (black line) above-barrier illumination with $P_\text{a} = 0.04$~W/cm$^2$. $B = 7$~T and $T = 1.8$~K. (a) On the exciton ($E_\text{R} = 1.6043$~eV) in Voigt geometry. (b) On the trions ($E_\text{R} = 1.6007$~eV) in tilted geometry ($\theta = 25^\circ$). Here, both electron- and heavy-hole-SFRS lines are visible for the T$^+$ resonance. (c) On the trions ($E_\text{R} = 1.6007$~eV) in Voigt geometry.}
\end{figure}

Further details on the electron and heavy-hole $g$ factor tensors as well as the SFRS processes are revealed by the dependences of the Raman shifts on the angle between magnetic field direction and $z$-axis. In Figure~\ref{fig6} the angular dependences of the electron- and heavy-hole-SFRS are shown for $B = 7$~T. The e-SFRS line (closed squares) weakly shifts with the angle $\theta$ increasing from the Faraday geometry ($\theta = 0^\circ$) to the Voigt geometry ($\theta = 90^\circ$). On the contrary, the hh-SFRS line (closed circles), measured at the exciton resonance, shows a strongly anisotropic behavior. The Raman shift can be described by the longitudinal and transverse $g$ factor values. The dependence of the $g$ factor on the tilting angle $\theta$ is given by
\begin{equation} \label{eqgfactor}
g(\theta) = \sqrt{(g^{\parallel} \cos \theta )^2+(g^{\perp} \sin \theta )^2} .
\end{equation}
Fits of the spin-flip Raman shifts $\Delta E_{\text{SF}}(\theta) = g(\theta) \mu_{\text{B}} B$ by Eq.~(\ref{eqgfactor}) give the following $g$ factor tensor components of electron and heavy-hole~\cite{notegfactor}:
\begin{eqnarray*}
g_{\text{e}}^{\parallel} = -1.71 \pm 0.01 \; &,& \; g_{\text{e}}^{\perp} = -1.62 \pm 0.01 ,\\
g_{\text{hh}}^{\parallel} = -1.52 \pm 0.01 \; &,& \; g_{\text{hh}}^{\perp} = -0.15 \pm 0.02 .
\end{eqnarray*}
Due to the nonlinear magnetic field dependence of the hh-SFRS line in  Faraday geometry, see Fig.~\ref{fig4}, the given $g_{\text{hh}}^{\parallel}$ value is valid for the particular field of $B=7$~T. For  $B < 6$~T the longitudinal $g$ factors of electron and heavy-hole are about equal. The conclusion on the negative sign of $g_{\text{hh}}^{\perp}$ is made from the fact that for all angles its value is larger than zero~\cite{Ilya}. From these data, the longitudinal and transversal $g$ factors of the exciton can be estimated. By use of the definition $g_{\text{X}} = g_{\text{hh}} - g_{\text{e}}$ for the bright exciton we obtain $g_{\text{X}}^{\parallel} = 0.19 \pm 0.01$ and $g_{\text{X}}^{\perp} = 1.47 \pm 0.02$ at $B=7$~T.

SFRS spectra for p-type and n-type regimes measured, respectively, without and with illumination are shown in Fig.~\ref{fig7} for different geometries and excitation conditions. In Voigt geometry and at the exciton resonance both e- and hh-SFRS lines are observed in the p-type regime, see the black curve in Fig.~\ref{fig7}(a). Due to the fact that the $g_{\text{hh}}^{\perp}$ value is very small the hh-SFRS line is seen as a shoulder in close vicinity to the laser line. Under illumination the hole signal disappears and the SFRS spectrum (red curve) solely contains the e-SFRS line. Note, in Voigt geometry the ratios between the e-SFRS line maxima and the PL background are smaller than the corresponding ratios in the Faraday geometry, compare with Figs.~\ref{fig2}(a) and \ref{fig2}(b).

The angular dependence of the SFRS probed at the positive trion manifests itself in a strong variation of the SFRS intensities. In Faraday geometry the T$^+$ excitation yields an e-SFRS line, marked by open diamonds in Fig.~\ref{fig6}, while a heavy-hole spin-flip is absent. In tilted geometry with an angle from $\theta \approx 15^\circ$ to $25^\circ$ the spin-flip of both electron and heavy-hole can be identified from their partially overlapping SFRS lines. A corresponding SFRS spectrum for $\theta = 25^\circ$ is shown by the black curve in Fig.~\ref{fig7}(b). The e-SFRS intensity decreases with increasing angle and vanishes for $\theta > 25^\circ$, while the hh-SFRS becomes stronger and demonstrates the same angle-dependent shift as the one at the exciton resonance, see open triangles in Fig.~\ref{fig6}. These angle dependences imply that the e- and hh-SFRS processes, probed at the T$^+$ trion, have different scattering probabilities and involve SFRS mechanisms, which are differently influenced by the tilting of the magnetic field. It is interesting that, while the e-SFRS at the T$^+$ remains disappeared in the Voigt geometry, it can be measured on the T$^-$ trion under illumination, as demonstrated in Fig.~\ref{fig7}(c).

\begin{figure}[t]
\centering
\includegraphics[width=6.9cm]{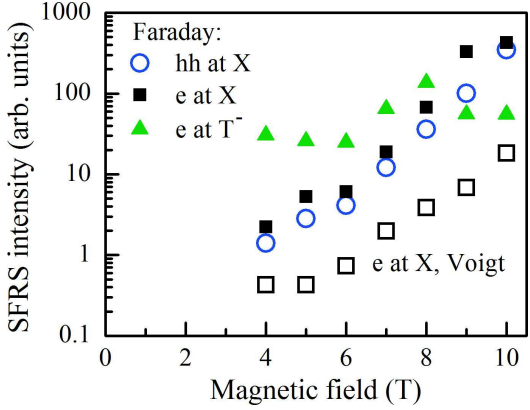}
\caption{\label{Bdep} (Color online) Intensities of the electron- and heavy-hole-SFRS lines shown as a function of magnetic field. They are measured for Faraday and Voigt geometries at the exciton and trion resonances, $T=1.8$~K. In Faraday geometry, the counter-circular for the e-SFRS and co-circular polarization for the hh-SFRS was chosen. In Voigt geometry, the polarization was set to crossed-linear.}
\end{figure}

The magnetic field dependences of the SFRS line intensities are shown in Fig.~\ref{Bdep}. In Faraday geometry, the linear evolutions in the half-logarithmic representation demonstrate that the e- and hh-SFRS intensities probed at the exciton resonance increase exponentially with $B$ by much more than two orders of magnitude from 4 to 10~T. On the contrary, the e-SFRS intensity increases by much less than two orders of magnitude in Voigt geometry. The e-SFRS intensity at the T$^-$ resonance is about constant with increasing magnetic field in Faraday geometry, see the green triangles.

\subsection{Temperature and power dependence of SFRS} \label{sectemp}
\begin{figure}[t]
\centering
\includegraphics[width=6.9cm]{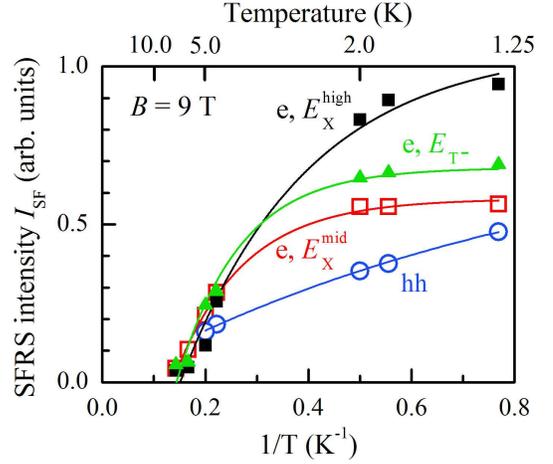}
\caption{\label{fig8} (Color online) Temperature dependence of the electron- and heavy-hole-SFRS intensities for resonant excitation of the exciton and T$^-$ trion (Faraday geometry). Exponential fits after Eq.~(\ref{tempexpo}), from which the deactivation energies of the SFRS processes are evaluated, are given by lines. The e-SFRS is probed in resonance with the exciton at $E_\text{X}^{\text{mid}}=1.6035$~eV or $E_\text{X}^{\text{high}} = 1.6043$~eV as well as the negative trion at $E_{\text{T}^-} = 1.6011$~eV, the polarization configuration is $(\sigma^+, \sigma^-)$. The hh-SFRS excited in the exciton is measured in $(\sigma^+, \sigma^+)$ configuration.}
\end{figure}

In Figure~\ref{fig8} the intensities of the electron- and heavy-hole-SFRS lines are shown as function of the lattice temperature varied from 1.3 up to 9~K (see top scale). The measurements were done in Faraday geometry at $B=9$~T. The SFRS processes induced in the exciton complex by excitation with a photon energy $E_\text{X}^{\text{mid}} = 1.6035$~eV are indicated by open symbols, while the closed squares refer to the e-SFRS at higher energy $E_\text{X}^{\text{high}} = 1.6043$~eV. The e- and hh-SFRS lines are strongly sensitive to temperature. Already at $T = 5$~K their intensities are at least twice smaller than that at 2~K, whereas at 9~K they almost vanish. The temperature-dependent SFRS intensities follow Arrhenius-like exponential equations
\begin{equation} \label{tempexpo}
I_\text{SF} = I_0 \left [ 1 - \exp \left ( - \frac{\epsilon}{k_\text{B} T} \right ) \right ] ,
\end{equation}
where $I_0$ is the intensity amplitude, $k_\text{B}$ the Boltzmann constant, and $\epsilon$ the deactivation energy of the corresponding SFRS process. The hh-SFRS is described by a deactivation energy of only $0.1$~meV. By comparison, the fittings of the e-SFRS data yield deactivation energies of $0.3$~meV for $E_\text{X}^{\text{high}}$ and $0.8$~meV for $E_\text{X}^{\text{mid}}$. Although the absolute intensity of the high-energy e-SFRS is large, its deactivation energy is about three times smaller than that for $E_\text{X}^{\text{mid}}$. The deactivation energies of the e-SFRS processes are smaller for larger excitation energies. Moreover, $\epsilon$ for $E_\text{X}^{\text{mid}}$ is similar to $\epsilon = 0.7$~meV for the negative trion resonance, see the green triangles in Fig.~\ref{fig8}.

Since the exciton binding energy in the studied CdTe/(Cd,Mg)Te QW is about 12~meV and that of a trion is about 2-3~meV,~\cite{Zhukovc,Sergeev} $\epsilon$ cannot be attributed to dissociation of either trions or excitons. The energies $\epsilon$ for the exciton and negative trion are rather comparable with the line width of their PL lines, which is about 0.9~meV, see Fig.~\ref{fig1}(b). These inhomogeneous PL line widths are determined by fluctuations of the confinement potential along the QW plane due to QW width and alloy fluctuations. Therefore, we associate the deactivation energy with the localization energy of the corresponding exciton complex or rather with the energy to excite a carrier from localized to quasi-free states. For higher excitation energy, the exciton complexes become less localized, leading also to a reduction in the deactivation energy. This underlines that localization is essential for SFRS processes to occur. For delocalized exciton complexes, however, the SFRS disappears.

\begin{figure}[t]
\centering
\includegraphics[width=6.9cm]{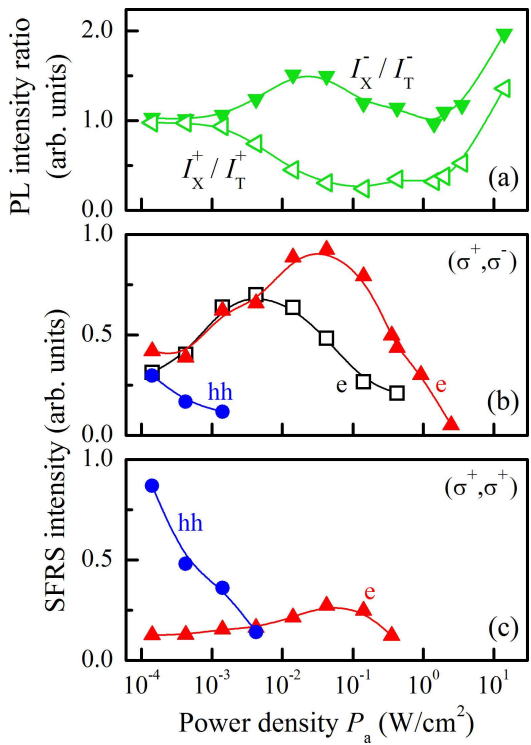}
\caption{\label{fig9} (Color online) Dependence of the SFRS and PL intensities on the density of the above-barrier illumination in Faraday geometry at $B=9$~T and $T=1.8$~K. (a) Intensity ratios of the $\sigma^+$ or $\sigma^-$ polarized PL between exciton (X) and trion (T), excited only by $E_{\text{a}} = 2.33$~eV photons. (b) and (c) Enhancement and quenching of the electron- and heavy-hole-SFRS intensities by the illumination in cross- $(\sigma^+, \sigma^-)$ and co-circularly $(\sigma^+, \sigma^+)$ polarized configurations. The density of the resonant excitation is $P_{\text{R}} = 0.5$~W/cm$^2$. The e-SFRS for resonant excitation of the trion ($E_\text{R} = 1.6011$~eV) is marked by open squares, the other experimental data (closed symbols) are taken at the exciton resonance ($E_\text{R} = 1.6037$~eV).}
\end{figure}

In the following the impact of the power density of above-barrier illumination on the SFRS processes will be highlighted. A gradual increase of the illumination density allows us to tune the resident carrier concentration in the QW and change its type from holes to electrons. Respective changes can be detected by PL spectra, and should be evident in SFRS spectra as well. Experimental information about the response of PL and SFRS signals on the illumination intensity, which changes by five orders of magnitude, is collected in Fig.~\ref{fig9}.

The turnover in the type of the resident carriers can be conveniently traced by the changes in the ratio of exciton and trion PL intensities, $I_\text{X}/I_\text{T}$, measured in two circular polarizations. One can see in Fig.~\ref{fig1}(d) that the magnetic field dependences of the circular polarization degree are similar for the exciton and T$^+$, while T$^-$ has a strongly different behavior. Therefore, one cannot expect changes in both polarizations for the $I_\text{X}/I_\text{T}$ dependences, when the resident holes prevail. Respectively, a strong dependence should appear when the resident electrons will take over with increasing illumination density. Such a turnover can be clearly seen in Fig.~\ref{fig9}(a) at $P_{\text{a}} = 2 \times 10^{-3}$~W/cm$^2$.

The presence of such turnover in the resident carrier type is nicely demonstrated by the SFRS intensities, which are shown in Figs.~\ref{fig9}(b) and \ref{fig9}(c). Here, different excitation densities and polarization configurations were chosen to highlight both electron and hole Raman signals. With increasing $P_{\text{a}}$ the hh-SFRS looses intensity (see closed circles) and disappears for $P_{\text{a}} > 4 \times 10^{-3}$~W/cm$^2$. At the same time the e-SFRS intensity increases and reaches maximum at $P_{\text{a}} \approx 5 \times 10^{-2}$~W/cm$^2$ for resonant excitation of the exciton (red triangles) and at about $5 \times 10^{-3}$~W/cm$^2$ for the T$^-$ resonance (open squares). At higher illumination densities the e-SFRS intensities decrease and vanish above 0.5~W/cm$^2$.

\begin{table*}
\begin{ruledtabular}
\caption{\label{taboverview} Properties of the different SFRS lines for resonant excitation of the exciton ($E_\text{X}$) and trion ($E_\text{T}$) complexes. The properties in columns 3 and 4 refer to Stokes scattering. Above-barrier illumination is indicated by $E_\text{a}$. The deactivation energy $\epsilon$ and the dominant circular polarization configuration ($\sigma^\eta , \sigma^\gamma$), abbreviated by ($\eta \, \gamma$), are designated. The geometries, in which the SFRS lines are observed, are marked by the tilting angle range. The main SFRS mechanisms for Faraday and tilted geometries are listed in columns 6 and 7. They refer to anisotropic exchange interaction ($\hat{H}_\text{exch}$), interaction with an acoustic phonon, fast hole spin relaxation ($1/\tau_s^\text{h} \gg \delta$), magnetic-field induced electron state mixing, and light-heavy-hole mixing.}
\begin{tabular}{ccccccc}
SFRS & Excitation & $\epsilon$ (meV) & ($\eta \, \gamma$) & Geometry & Mechanism $\theta = 0^\circ$ & Mechanism $\theta \not = 0^\circ$ \\ \hline
e & $E_\text{X}$ (+$E_\text{a}$) & 0.8 & ($+-$)$>$($++$) & $0^\circ \leq \theta \leq 90^\circ$ & phonon $+$ $1/\tau_s^\text{h} \gg \delta$, $\hat{H}_{\text{exch}}^\text{e-h}$ & phonon $+$ $1/\tau_s^\text{h} \gg \delta$, $\hat{H}_{\text{exch}}^\text{e-h}$ \\
hh & $E_\text{X}$ (+$E_\text{a}$) & 0.1 & ($++$)$>$($+-$) & $0^\circ \leq \theta \leq 90^\circ$ & $\hat H_{\text{exch}}^{\text{h-rh}}$ & $\hat H_{\text{exch}}^{\text{h-rh}}$ \\ \hline
hh & $E_\text{T}$ [T$^+$] & $\approx 0.3$ & crossed-lin.& $15^\circ \leq \theta < 90^\circ$ & & mixed e-states + $\text{lh}\leftrightarrow \text{hh}$ \\
e & $E_\text{T} + E_\text{a}$ [T$^-$] & 0.7 & ($+-$),($-\,-$) & $0^\circ \leq \theta \leq 90^\circ$ & $\hat{H}_{\text{exch}}^\text{e-hh}$ & mixed e-states + $\text{lh}\leftrightarrow \text{hh}$ \\
\end{tabular}
\end{ruledtabular}
\end{table*}

The different positions of the e-SFRS intensity maxima for X and T$^-$ excitation and their differently strong decreases with increasing power density indicate that they differ in their physical origins. For the exciton, its lifetime is shortened at high illumination intensities due to the increased probability of being captured to T$^-$ trion. At the T$^-$ resonance, heating of the resident electrons and their delocalization are relevant. They reduce the lifetime of the scattering state and, in turn, the intensity of the e-SFRS line. Such explanation is supported by the strong temperature dependence shown in Fig.~\ref{fig8} and the nonmonotonic changes in PL intensities at higher illumination densities $P_{\text{a}} > 1$~W/cm$^2$ in Fig.~\ref{fig9}(a).

In order to demonstrate that the carrier separation (electrons are collected into the QW and holes are captured in the barriers) under above-barrier illumination is the leading mechanism of the observed changes in SFRS signals, we performed experiments at below-barrier illumination with an energy of $1.88$~eV. It was found that for similar power densities as used for the above-barrier illumination in Fig.~\ref{fig9} the SFRS intensity remains unchanged. Only at larger illumination densities the SFRS lines become suppressed, most probably due to delocalization of the resident carriers or shortening of the exciton lifetime.

In Table~\ref{taboverview} the characteristics of the experimentally observed SFRS lines are summarized. The type of the SFRS process is classified with respect to the excitation energy, which is in resonance with either exciton or trion. Additionally, the application of above-barrier illumination is indicated. The deactivation energy as well as the dominant circular polarization configuration are specified. The mechanisms responsible for the SFRS processes in Faraday and tilted geometries, which will be discussed in detail in the following sections, are also given in the Table.

\section{Model consideration} \label{discussion}
\subsection{Faraday geometry}
Insight into the SFRS mechanism can be gained by analyzing the underlying selection rules for the circularly polarized Raman spectra measured in Faraday and backscattering geometry. In the electric-dipole approximation the total change of the photon angular momentum projection amounts to $\Delta l = 0$ or $\pm 2$. When the projection of the total angular momentum of the exciton complex $j_z$ on the $z$-quantization axis changes by $\pm 2$ the transitions are circularly co-polarized, while for $\Delta l = 0$ they are either linearly or crossed-circularly polarized. For resonant probing of a heavy-hole exciton, the optical excitation of the states $| j_z \rangle = |s_{z,\text{e}},j_{z,\text{hh}} \rangle = | +1 \rangle = |- 1/2 , + 3/2 \rangle$ and $| -1 \rangle = |+ 1/2 , - 3/2 \rangle$ by absorption of $\sigma^+$ and $\sigma^-$ polarized light, respectively, is allowed. Here, $s_{z,\text{e}}$ and $j_{z,\text{hh}}$ designate the projections of the electron spin and heavy-hole angular momentum onto the $z$-axis.

In the following the electric-dipole selection rules are analyzed for SFRS processes involving a neutral exciton, negative or positive trion as an intermediate state. For an ideal (100)-grown CdTe/(Cd,Mg)Te QW the symmetry is given by the irreducible representation $D_{2d}$. The heavy-hole subbands are separated from the light-hole ones by quantum confinement and strain caused by mismatch of the QW and barrier lattice constants. A magnetic field applied along the QW growth axis lifts the spin degeneracy of the electron and heavy-hole states. For simplicity we consider the Stokes scattering of a $\sigma^+$ polarized incident photon. The possible SFRS processes for the three exciton complexes are schematically illustrated in Fig.~\ref{fig10}:

\begin{itemize}
\item[(i)] \textit{Exciton}

The incident $\sigma^+$ polarized photon creates an exciton in the state $| +1 \rangle = |- 1/2 , + 3/2 \rangle$. A single spin-flip of either the electron or heavy-hole leads to a dark exciton state of $|+2 \rangle = |+ 1/2 , + 3/2 \rangle$ or $|-2\rangle = |- 1/2 , - 3/2 \rangle$, respectively. Hence, Raman scattering of a single carrier spin in an exciton is not observable. However, the simultaneous reversal of the electron and heavy-hole spins is allowed. This exciton spin-flip scattering process from $|+1\rangle$ to $|- 1 \rangle$ is induced by an acoustic phonon, as shown in Fig.~\ref{fig10}(i). The exciton annihilation yields a scattered photon having opposite circular polarization than the incident one, i.e. $\sigma^-$ polarization in the considered example. Therefore, the exciton-SFRS is observable in the $(\sigma^\eta, \sigma^{-\eta})$ configuration. The Raman shift is given by the exciton Zeeman splitting: $\Delta E_\text{SF} = \Delta E_\text{X}$.

\begin{figure*}[t]
\centering
\includegraphics[width=14.2cm]{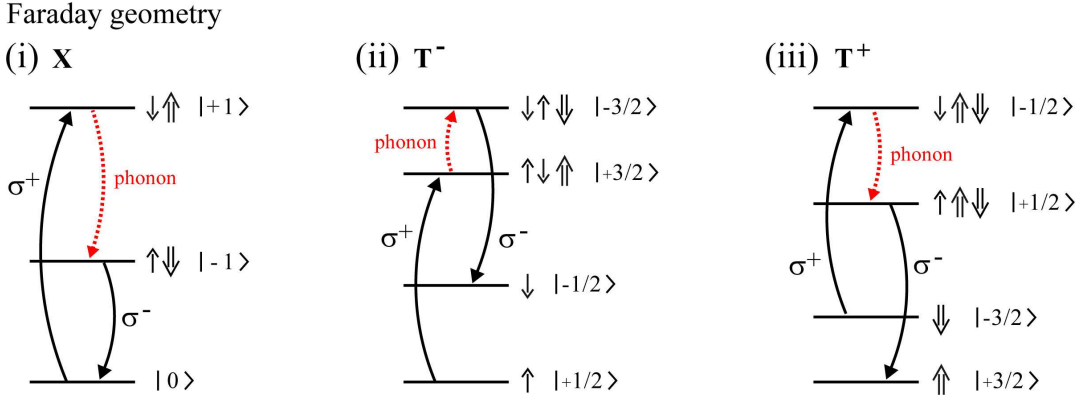}
\caption{\label{fig10} Schemes of electric-dipole allowed SFRS Stokes-processes for exciton, negative and positive trions. The Faraday geometry is considered for uncoupled light-hole and heavy-hole states. The resonant excitation is $\sigma^+$ polarized. (i) For an exciton the flip of its spin is allowed via acoustic phonon scattering. (ii) In the T$^-$ the heavy-hole spin is scattered by an acoustic phonon. The difference between the incident and scattered photon energies is $\hbar \omega_\text{i} - \hbar \omega_\text{s} > 0$ thus representing a Stokes process. (iii) In T$^+$ the unpaired electron spin interacts with an acoustic phonon. For both trions the Raman shifts are equal to $\Delta E_\text{e} + \Delta E_\text{hh}$.}
\end{figure*}
\item[(ii)] \textit{Negative trion, singlet state}

The $\sigma^+$ polarized incident photon can excite a negative singlet trion $| +1/2, -1/2, + 3/2\rangle$. Hereby, a resident electron with spin $+1/2$ should be involved, see Fig.~\ref{fig10}(ii). For T$^-$ the electron spin-flip (SF) is forbidden by Pauli's exclusion principle. Only the unpaired heavy-hole can flip its spin via acoustic phonon scattering, thus, the T$^-$ trion is scattered from $|+1/2, -1/2, +3/2 \rangle$ to $|+1/2, -1/2, -3/2 \rangle$. The T$^-$ annihilation leaves behind a $\sigma^-$ polarized scattered photon and a spin-down electron. Therefore, the relevant polarization configuration is $(\sigma^\eta,\sigma^{-\eta})$, and the Raman shift is equal to $\Delta E_\text{SF} = \Delta E_\text{e} + \Delta E_\text{hh}$, coinciding with that of the exciton. Note, since the sign of $\Delta E_\text{e(hh)}$ depends on the sign of the electron (heavy-hole) $g$ factor, $\Delta E_\text{e(hh)}$ can be either positive or negative.

\item[(iii)] \textit{Positive trion, singlet state}

By analogy with the negative trion the resonant probing of the positive trion $| -1/2, -3/2, +3/2 \rangle$ solely yields a SFRS process of the unpaired electron spin. It scatters from $s_{z,\text{e}} = -1/2$ to $+1/2$ via an acoustic phonon, see Fig.~\ref{fig10}(iii). The annihilation of the $| +1/2, -3/2, +3/2\rangle$ trion results in a $\sigma^-$ polarized photon and leaves a $|+3/2 \rangle$ heavy-hole. The respective selection rule is $(\sigma^\eta, \sigma^{-\eta})$, and the Raman shift is again $\Delta E_\text{SF} = \Delta E_\text{e} + \Delta E_\text{hh}$.
\end{itemize}

One can figure out that all three processes shown in Fig.~\ref{fig10} have several common features.
\begin{itemize}
\item Second-order Raman process, which only accounts for incoming (absorption) and outgoing (emission) photons, is forbidden~\cite{Cardonabook}. Instead, the SFRS processes involve an intermediate state, thus, they are of third order. 
\item Scattering is provided by an acoustic phonon.
\item The scattered photon has opposite polarization with respect to the incident photon.
\item For spin-flip processes within trions the unpaired carrier spin is actively flipped, the resident carrier is left with opposite spin orientation.
\item All processes have the same Raman shift $\Delta E_\text{SF} = \Delta E_\text{X}= \Delta E_\text{e} + \Delta E_\text{hh}$.
\end{itemize}

\subsection{Voigt geometry}
\begin{figure*}[t]
\centering
\includegraphics[width=14.9cm]{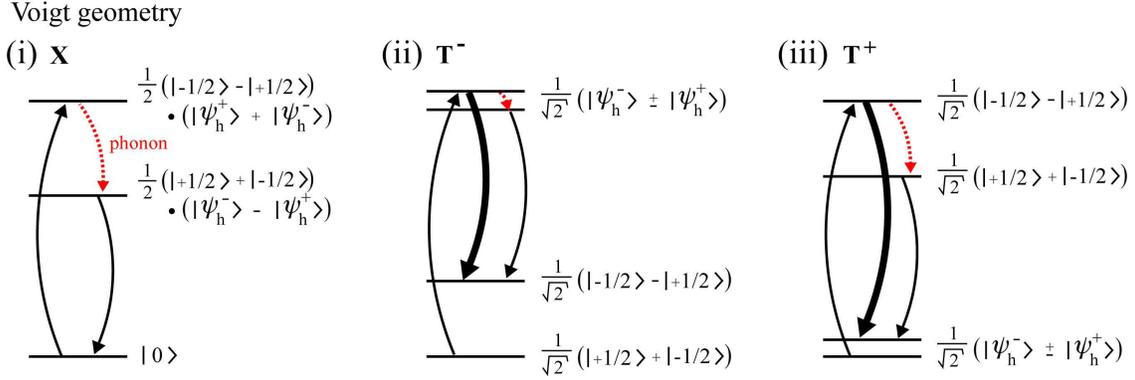}
\caption{\label{fig10c} Schemes of electric-dipole allowed SFRS Stokes-processes for exciton, negative and positive trions in Voigt geometry. The resonant excitation is linearly polarized. Second-order Raman processes are shown by thick arrows in the scattering channel, and thin arrows indicate third-order processes involving phonons. (i) The exciton spin is flipped via acoustic phonon scattering. (ii) For T$^-$ a second-order process provides an e-SF. The Raman shift for the third-order process is equal to $\Delta E_\text{e} + \Delta E_\text{hh}$. (iii) For T$^+$ a second-order process provides a hh-SF, which is rather small, and the shift for the third-order one is given by $\Delta E_\text{e} + \Delta E_\text{hh}$. Note, for both trions in the singlet state only the wave functions of the unpaired carrier are designated.}
\end{figure*}

In semiconductor QWs with cubic symmetry the spin-flip Raman scattering is governed by the differential scattering cross-section being proportional to $| (\mathbf{\hat e}_{\text{i}} \times \mathbf{\hat e}_{\text{s}}^*) \times \mathbf{\hat B} |^2$. In the backscattering geometry, the cross product of the polarization vectors of the incident $\mathbf{\hat e}_{\text{i}}$ and scattered light $\mathbf{\hat e}_{\text{s}}$ with the magnetic field vector $\mathbf{\hat B}$ differs from zero only if the magnetic field has a nonzero component perpendicular to the light propagation direction. Contrary to the Faraday geometry, in the Voigt geometry the external magnetic field is perpendicular to the light wave vector and to the structure $z$-axis. In this case the carriers and exciton complexes cannot be described by spin basis states. As a result, for trions second-order Raman processes become allowed. They do not require intermediate phonon scattering and, therefore, have a higher probability in comparison to the third-order processes considered for the Faraday geometry. Nevertheless, the third-order processes are also allowed in Voigt geometry as additional ones for trions and as the main process for the exciton. These details are schematically illustrated in Fig.~\ref{fig10c}. Hereby, in order to reflect the real properties of the studied structure with nonzero value of $g_{\text{hh}}^{\perp}$, we consider the heavy-hole wave functions on account of finite admixture of the light-hole states~\cite{Koudinovc,Leger}. The optical transitions are analyzed in commonly used crossed-linear polarizations.

As indicated by the nonlinear magnetic field dependence of the heavy-hole-SFRS shift, see Fig.~\ref{fig4}, light-hole and heavy-hole states become mixed. For valence-band mixing induced by local QW deformations or an asymmetric interface potential~\cite{Koudinovc,Leger,Pikus,Krizhanovskii}, the lowest energy hole states take the form
\begin{equation} \label{eqexchangeeh}
|\mathit{\Psi}_{\text{h}}^{\pm} \rangle \propto |\pm 3/2 \rangle - \beta |\mp 1/2 \rangle .
\end{equation}
Here, $\beta$ characterizes the hh-lh admixture. The hole wave functions in Voigt geometry are superpositions given by
\begin{equation*} \label{holetotalvoigt}
\frac{1}{\sqrt{2}} \left ( |\mathit{\Psi}_\text{h}^{-} \rangle \pm |\mathit{\Psi}_\text{h}^{+} \rangle \right ) .
\end{equation*}
The electron wave functions are also composed of both spin basis states with equal probability:
\begin{equation*} \label{electrontotalvoigt}
\frac{1}{\sqrt{2}} \left ( |\pm 1/2 \rangle \pm |\mp 1/2 \rangle \right ) .
\end{equation*}
The confined bright and dark exciton states can be factorized into the product of the electron and hole states.

It is worthwhile to note that in the case when the mixing of hh- and lh-states in the QW is absent, $g_{\text{hh}}^{\perp}=0$ and the hole state does not split in the Voigt geometry. As a result, the second-order SFRS for the T$^+$ trion is not detectable, as the scattered photon energy coincides with the laser excitation energy. Also in this case, the Raman shifts provided by third-order processes for exciton and trions coincide with the electron Zeeman splitting, since $\Delta E_\text{hh}$ vanishes.

\subsection{Tilted geometry}
In tilted geometry the contribution of the spin basis states to the specific spin state is additionally controlled by the tilting angle $\theta$ between the magnetic field direction and QW growth axis. The mixing of the electron states established via the coupling between the in-plane component of the magnetic field vector and transverse electron $g$ factor can be described by~\cite{Thomas}:
\begin{equation} \label{eqwavefunceltilt}
|\mathit{\Psi}_{\text{e}}^{\pm} \rangle = \cos(\theta/2) | \pm 1/2 \rangle \pm \sin(\theta/2) | \mp 1/2 \rangle .
\end{equation}

The hh-lh mixing is affected by the coupling of $\mathbf{B}$ to a nonzero in-plane magnetic moment of the heavy-hole due its transverse $g$ factor $g_\text{hh}^\perp \not = 0$. Therefore, in tilted geometry the probability of the mixing between the heavy-hole and light-hole states differs from that in Faraday and Voigt geometry. Correspondingly, the hh-lh mixing coefficient $\beta$ depends on the tilting angle $\theta$. The hole wave functions take the following form:
\begin{equation} \label{holewavetilted}
|\mathit{\Phi}_{\text{h}}^{\pm} \rangle = \cos(\theta/2) |\mathit{\Psi}_{\text{h}}^{\pm} \rangle \pm \sin(\theta/2) |\mathit{\Psi}_{\text{h}}^{\mp} \rangle .
\end{equation}

Taking the T$^+$ trion as example, incident light which is vertical-linearly polarized excites the positive trion into the state $|\mathit{\Psi}_{\text{e}}^{-} , \mathit{\Phi}_{\text{h}}^{-}, \mathit{\Phi}_{\text{h}}^{+} \rangle$, whereby the resident hole initially occupies the $| \mathit{\Phi}_{\text{h}}^{+} \rangle$ state. Due to the mixed conduction band states, the axial symmetry of the complex, and the occupation of both hole spin states, neither the electron nor the holes need to change their spin states to allow for a hh-SFRS process. The final stage of this second-order SFRS process is governed by the annihilation of the trion $|\mathit{\Psi}_{\text{e}}^{-} , \mathit{\Phi}_{\text{h}}^{-}, \mathit{\Phi}_{\text{h}}^{+} \rangle$ which leaves a hole in the $| \mathit{\Phi}_{\text{h}}^{-} \rangle$ state. The energy difference between the incident and scattered light is determined by the hole Zeeman splitting.

\begin{figure*}[t]
\centering
\includegraphics[width=12.8cm]{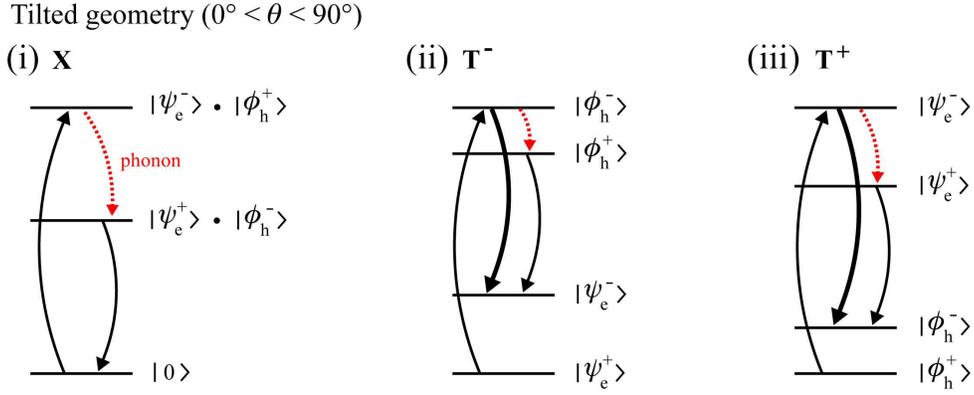}
\caption{\label{fig10b} Schemes of electric-dipole allowed SFRS Stokes-processes for exciton, negative and positive trions in tilted geometry. The SFRS mechanisms correspond to that in the Voigt geometry, only the hole spin-level separation is larger. Note, for both trions in the singlet state only the wave functions of the unpaired carrier are designated.}
\end{figure*}

The respective Raman processes allowed in second and third orders are schematically shown in Fig.~\ref{fig10b}. One can easily see by comparing with Fig.~\ref{fig10c} that the processes allowed in tilted and Voigt geometries are similar. The main difference results from the larger Zeeman splitting of the heavy-hole in the tilted case, so that the hh-SF becomes well separated from the laser line and can be detected experimentally.

\subsection{Carrier-carrier exchange interaction} \label{exch_int}
Exchange interaction between carriers is a key ingredient of the spin-flip related Raman scattering. Two types of exchange interactions between two spin carriers in a QW can lead to spin flip-flop or flip-stop Raman processes.

The isotropic carrier-carrier exchange provides the flip-flop of both carrier spins. Considering Faraday geometry ($\mathbf{B} \parallel \mathbf{z}$), the spin flip-flop is described by the Heisenberg-like Hamiltonian
\begin{equation} \label{eqflipflopdirect}
\hat{H}_{\text{d}} = J_{\text{d}} \, \left ( \sigma_{x,1} \sigma_{x,2} + \sigma_{y,1} \sigma_{y,2} \right ) .
\end{equation}
Here, $J_{\text{d}}$ is the exchange integral and $\sigma_{x,y}$ are the Pauli matrices of the two carriers.

Anisotropic flip-stop-like exchange interaction results from spatially shifted probability densities of two spin carriers; in very common case we consider electron and hole of an exciton. The centers $\boldsymbol{\rho}_\text{e}$ and $\boldsymbol{\rho}_\text{h}$ of their in-plane localization areas shall not coincide. The presence of the in-plane preferential direction $\boldsymbol{\rho}_\text{e} - \boldsymbol{\rho}_\text{h}$ lowers the symmetry of the exciton complex. Since the projection of the angular momentum on the axis defined by $\boldsymbol{\rho}_\text{e} - \boldsymbol{\rho}_\text{h}$ is not preserved, the restrictions imposed by the angular momentum conservation are lifted. This symmetry reduction allows an anisotropic exchange interaction which flips the spin of one carrier while leaving that of the other unchanged. Correspondingly, it is called flip-stop-like exchange interaction~\cite{Sapega,Sapegab,Ivchenko}. The Hamiltonian of the flip-stop-like e-h exchange interaction is given by
\begin{equation} \label{ehhexch}
\hat H_{\text{exch}}^{\text{e-h}} = (\Delta_\text{h} \sigma^{\text{h}}_{+} + \Delta_\text{h}^{*} \sigma^{\text{h}}_{-}) \sigma^{\text{e}}_z .
\end{equation}
Here, $\Delta_\text{h}$ and $\Delta^*_\text{h}$ are complex coefficients, and the Pauli matrices $\sigma^{\text{e}}_z$ and $\sigma^{\text{h}}_{\pm} = (\sigma_x \pm \text{i} \sigma_y)/2$ act on the electron or, respectively, light-hole and heavy-hole states which are described by their basis states $1/\sqrt{2} |x-\text{i}y\rangle |\downarrow \rangle$ and $-1/\sqrt{2} |x+\text{i}y \rangle |\uparrow \rangle$ for hh, $1/\sqrt{6} (|x-\text{i}y\rangle |\uparrow \rangle + 2|z\rangle |\downarrow \rangle)$ and $-1/\sqrt{6} (|x+\text{i}y\rangle |\downarrow \rangle - 2|z\rangle |\uparrow \rangle)$ for lh.

As indicated, the anisotropic exchange requires a symmetry reduction due to spatially mismatched probability densities of the in-plane electron and hole wave functions. To describe anisotropic exchange interaction between carriers with identical charges like e-e or h-h one can use the Hamiltonian Eq.~(\ref{ehhexch}) with respective Pauli matrices. In a magnetic field the Hamiltonian Eq.~(\ref{ehhexch}) is responsible for the spin-flip. The effective field which is produced by the one carrier due to its Larmor precession around the magnetic field direction is felt by the other carrier, and vice versa. The efficiency of e-h anisotropic exchange in a magnetic field is determined by the difference in the electron and hole $g$ factor values. If their $g$ factors are identical, the time-dependent probability of spin exchange will be proportional to $\sin^2(J_\text{exch} t/2)$, where $J_\text{exch}$ is the exchange constant~\cite{Burshtein}. For different $g$ factors the strength of the anisotropic exchange depends on the mismatch between the electron and hole $g$ factors. Thus, the probability of an exchange interaction process is scaled by the Larmor frequencies $\omega_\text{\tiny L} = g \mu_\text{B} B/\hbar$ of the electron and hole. The probability $p_\text{exch}$ for the anisotropic e-h exchange interaction is then given by~\cite{Molin}
\begin{equation} \label{intvsangl}
p_\text{exch} \approx \hbar^2/(\mu_\text{B} B)^2 \, \frac{J_\text{exch}^2}{\left ( \omega_{\text{\tiny L}}^{\text{e}} - \omega_{\text{\tiny L}}^{\text{h}} \right )^2 + \Gamma^2 } ,
\end{equation}
where $\Gamma$ is the damping constant being inversely proportional to the exciton lifetime. Consequently, for strongly anisotropic hole and isotropic electron $g$ factors one can expect a strong angular dependence of the Raman scattering efficiency for the flip-stop process.

\section{Discussion} \label{discusschap}
In this section, the model expectations considered in the previous section will be compared with the experimental data for the studied CdTe/(Cd,Mg)Te QW. We start our consideration with trions, where the situation is simpler and mostly in agreement with the model expectations. Then we move to the SFRS on the exciton, which is more complicated, and its explanation requires the treatment of additional mechanisms.

Among all the present experiments we did not observe a SFRS process that corresponds to $\Delta E_\text{SF} = \Delta E_\text{X}= \Delta E_\text{e} + \Delta E_\text{hh}$, as illustrated for Faraday geometry in Fig.~\ref{fig10}. But we found Raman shifts of $\Delta E_\text{e}$ and $\Delta E_\text{hh}$. Clearly, we need some new (different) mechanism to explain them. Obviously, symmetry reduction has to play an important role in this context.

Before we start to consider individual cases for exciton complexes a general comment on the specifics of the studied QW is helpful. As the electron and heavy-hole Zeeman splittings coincide in Faraday geometry in low magnetic fields and differ only slightly in higher fields, see Fig.~\ref{fig4}, the experimental observation of the SF line with $\Delta E_\text{SF} = \Delta E_\text{e} + \Delta E_\text{hh}$ is hindered by the laser stray light. Therefore, the observation of the SF process with $\Delta E_\text{e} + \Delta E_\text{hh}$ predicted for the exciton and trions by the schemes in Fig.~\ref{fig10} is difficult, and in fact was not found, see Figs.~\ref{fig2} and \ref{fig3}. But we could not judge from our data that these processes are not efficient. The exciton-SFRS was reported in earlier work on thinner CdTe/(Cd,Mg)Te QWs, where the electron and heavy-hole $g$ factors are not equal to each other~\cite{Sirenko}.

\subsection{SFRS at positive trion} \label{positive_trion}
Here, we discuss experimental results measured in the p-type regime, i.e., without above-barrier illumination, under resonant excitation of the T$^+$ trion state. In accordance with scheme (iii) of Fig.~\ref{fig10} carrier spin-flips are not present in Faraday geometry, see Fig.~\ref{fig3}(a). The weak e-SF feature in $(\sigma^+, \sigma^-)$ polarization is related to the minor presence of resident electrons. It has much stronger intensity for T$^-$ in the n-type regime, see Fig.~\ref{fig3}(b) and Sec.~\ref{negative_trion}. The SF based on a phonon-assisted Raman process is not seen, most probably due to the small value of $\Delta E_\text{e} + \Delta E_\text{hh}$, as discussed above.

In Voigt geometry shown in Fig.~\ref{fig7}(c) no SFRS features are found. The possible hh-SF suggested in Fig.~\ref{fig10c}(iii) is located too close to the laser line due to the small transverse hh $g$ factor. Owing to the strong background PL it cannot be resolved, in contrast to the resonant exciton excitation, see Fig.~\ref{fig7}(a), where the hh-SF is seen as a shoulder of the laser line.

In tilted magnetic fields the hh-SF is observed for the angles $15^\circ \lesssim \theta < 90^\circ$, see Figs.~\ref{fig7}(b) and 6 (open triangles). As noted in Table~\ref{taboverview}, this is in accordance with the dominant second-order process from Fig.~\ref{fig10b}(iii) based on mixed electron states. By analogy with the e-SF in Faraday geometry, the weak e-SF, shown in Fig.~\ref{fig7}(b) for $\theta = 25^\circ$, is due to the presence of resident electrons in the p-type regime, i.e., it is not attributed to the properties of the T$^+$ state.

\subsection{SFRS at negative trion} \label{negative_trion}
For T$^-$ in the n-type regime one can expect similar scenarios as for the T$^+$ trion. And this is indeed the case for Voigt and tilted geometries, see Figs.~\ref{fig7}(c) and \ref{fig7}(b), where the e-SF allowed in second-order is clearly seen. The strongest intensity is achieved in the Voigt geometry in agreement with the schemes of Figs.~\ref{fig10c}(ii) and \ref{fig10b}(ii).

\begin{figure}[t]
\centering
\includegraphics[width=6.0cm]{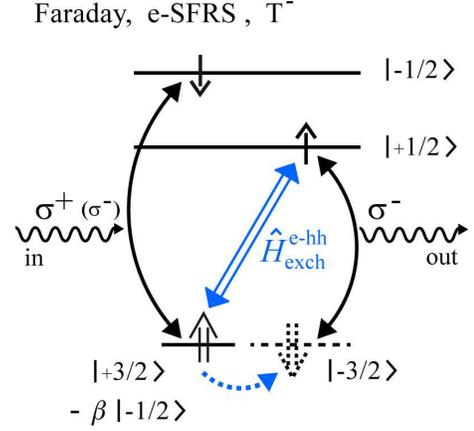}
\caption{\label{fig12} Scheme of e-SFRS Stokes-process for the negative trion in the Faraday geometry presented in the single-particle picture. It requires the anisotropic e-hh exchange interaction. The photo-hole is scattered into the virtual heavy-hole state $|-3/2\rangle$. The e-SF line is present in the $(\sigma^+, \sigma^-)$ and $(\sigma^-,\sigma^-)$ configurations. The latter is weak being proportional to the admixture of the lh state to the hh one. A radiative transition is marked by a curved arrowed single-line. The blue-colored arrowed double-line denotes anisotropic exchange interaction.}
\end{figure}

The surprising fact is the observation of the relatively strong e-SF in the $(\sigma^+, \sigma^-)$ polarization in Faraday geometry and the much weaker line in the $(\sigma^-, \sigma^-)$ polarization, see Fig.~\ref{fig3}(b) and Table~\ref{taboverview}. This cannot be explained within the frame of the process depicted in Fig.~\ref{fig10}(ii). It requires the involvement of an other mechanism. Such mechanism is suggested in Fig.~\ref{fig12}. Note, for convenience we choose the single-particle representation in the scheme, in contrast to the exciton-complex representation used in Figs.~\ref{fig10}, \ref{fig10c} and \ref{fig10b}. The resident electron is in the $|+1/2 \rangle$ state, and a $\sigma^+$ polarized photon generates a $|-1/2, +3/2 \rangle$ pair of carriers which forms a T$^-$ with the resident electron. Scattered light with $\sigma^-$ polarization results from the recombination of the $|+1/2 \rangle$ electron with a hole occupying the virtual state $|-3/2 \rangle$ in which the hole is scattered by anisotropic e-hh exchange interaction defined by Hamiltonian Eq.~(\ref{ehhexch}). The virtual hole state $|-3/2 \rangle$ has similar energy as the $|+3/2 \rangle$ state, and the hole spin splitting does not contribute to the Raman shift. The suggested mechanism is of third-order and can be relatively efficient. The presence of the weak signal in $(\sigma^-,\sigma^-)$ polarization configuration is due to the admixture of lh into hh state.

Two alternative mechanisms can be proposed. In a fully symmetric negative trion where the electron spins form a singlet, an e-SF will only be observable if the hole spin relaxes very fast~\cite{dyakonov1997}. However, one can expect that the hole spin relaxation in T$^-$ is slow~\cite{Zhukov}. Also, a SFRS mechanism based on the admixture of triplet trion states to the singlet ones can be excluded for magnetic fields smaller than 20~T,\cite{Bartsch} since at low $B$-fields they are separated in energy by more than 2~meV.

\subsection{SFRS at exciton resonance} \label{disexciton}
The exciton-SF as suggested by Figs.~\ref{fig10}(i), \ref{fig10c}(i) and \ref{fig10b}(i) is not found in any geometries. In Faraday geometry, the exciton Zeeman splitting in the studied QW is too small to be resolved. In tilted and Voigt geometries, the absence of the exciton-SF line can be explained by its relative weakness and broadening, as well as by the presence of efficient alternative channels (mechanisms). However, the heavy-hole- and electron-SF lines appear in the Raman spectra shown in Figs.~\ref{fig2} and \ref{fig7}(a). In the following, we discuss possible scattering mechanisms for these processes by distinguishing between the p-type and n-type regimes.

\textbf{p-type, hh-SF}. In the p-type regime in Faraday geometry a strong hh-SF is observed in ($\sigma^+,\sigma^+$) polarization, see Fig.~\ref{fig2}(a). In the three other polarizations its intensity is about 30\% of the highest one. This line shows a strong anisotropic behavior (Fig.~\ref{fig6}), as expected for the hh spin in a QW, and is even resolved in Voigt geometry, see Fig.~\ref{fig7}(a), due to the finite value of $g_{\text{hh}}^{\perp}$. The hh-SF line disappears with increasing illumination intensity, i.e., with the transition to the n-type regime, demonstrated in Fig.~\ref{fig9}(c).

\begin{figure}[t]
\centering
\includegraphics[width=7.9cm]{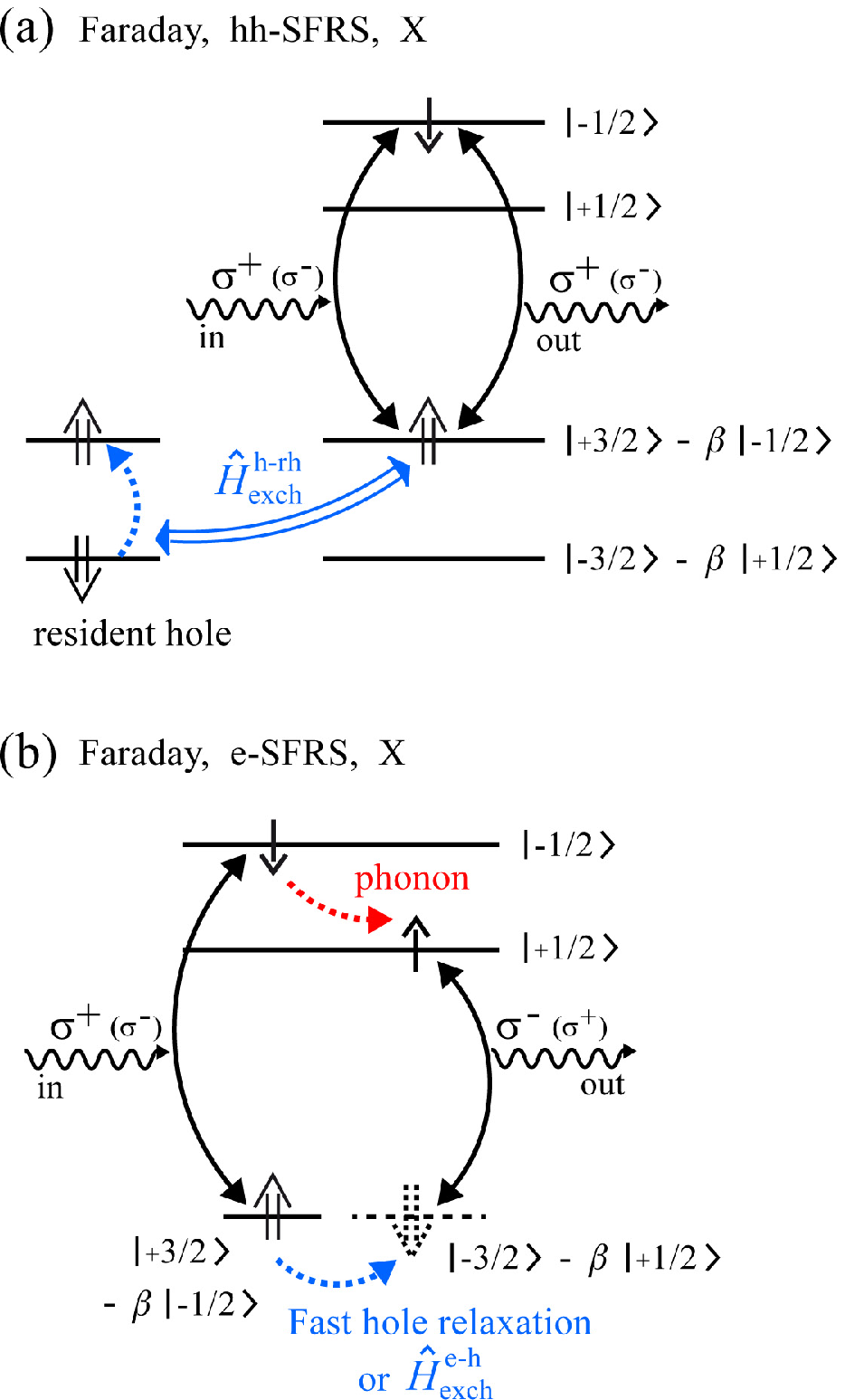}
\caption{\label{fig15} Schemes of the hh- and e-SFRS Stokes-processes at the exciton resonance in Faraday geometry in the single-particle representation. (a) Third-order process describing the hh-SF for the different polarization configurations. The intermediate state is provided by flip-stop spin scattering of the resident hole by the photo-hole. (b) The e-SFRS process is based on interaction of the electron with an acoustic phonon at first intermediate stage. At second stage, the hole change its spin state due to its fast spin relaxation or anisotropic exchange between the electron and hole. The circular polarization degree of the incident and scattered light is indicated by the size of the $\sigma$-symbol.}
\end{figure}

The related third-order scattering process based on flip-stop hole exchange interaction is shown in Fig.~\ref{fig15}(a). In the case of Stokes scattering a circularly polarized photon creates a $|-1/2 , \mathit{\Psi}_\text{h}^+ \rangle$ exciton. A resident hole (rh) is assumed to be neighbored to the photogenerated exciton. The wave function of the localized resident hole is spatially shifted from the wave function of the photo-hole. Due to the spatial in-plane separation the anisotropic exchange interaction between both holes becomes possible. The photo-hole induces a spin-flip of the resident hole from $|\mathit{\Psi}_\text{rh}^- \rangle$ to $|\mathit{\Psi}_\text{rh}^+ \rangle$ via the flip-stop process. By analogy with Eq.~(\ref{ehhexch}), the interaction Hamiltonian is given by $\hat H_{\text{exch}}^{\text{h-rh}} = (\Delta_\text{rh} \sigma^{\text{rh}}_{+} + \Delta_\text{rh}^{*} \sigma^{\text{rh}}_{-}) \sigma^{\text{h}}_z$. Finally, the $|-1/2, \mathit{\Psi}_\text{h}^+ \rangle$ exciton annihilates under emission of light having predominantly the same circular polarization as the incident one. The energy of the scattered light is reduced by the Zeeman energy of the resident hole ($\Delta E_\text{SF} = \Delta E_\text{hh}$). Due to the mixing of hole states it is partially polarized and is visible in other polarization configurations. The scattering mechanism is active in all geometries, as noted in Table~\ref{taboverview}. In tilted geometries, only the light polarization characteristics are changed particularly by the mixing of the electron states.

\textbf{n-type, e-SF}. In the n-type regime in Faraday geometry the e-SF is dominant in the ($\sigma^+, \sigma^-$) configuration and is present with $0.15$-$0.3$ intensity in the other polarization configurations, see Fig.~\ref{fig2}(b). This process is not expected for the exciton state, as one can see from Figs.~\ref{fig10}(i), \ref{fig10c}(i) and \ref{fig10b}(i), where the SF with $\Delta E_\text{X} =\Delta E_\text{e} + \Delta E_\text{hh}$ is allowed only. Before discussing the responsible mechanisms, let us summarize all properties of the e-SFRS under resonance excitation of the exciton in the n-type regime. Note, a weak e-SF line is also present in the spectrum for the p-type regime. It has all features of the e-SF in the n-type regime, and we explain it by the coexistence of resident electrons in the p-type regime.

\begin{itemize}
\item[(i)] The e-SF intensity in the Faraday geometry exceeds the one in Voigt geometry by a factor of 35. In order to explain this unexpected behavior only one mechanism related to magnetic localization comes into consideration. In high longitudinal magnetic fields the exciton wave function is shrunk, thus enhancing its localization and oscillator strength. This in turn leads to longer exciton dephasing and life times and, therefore, enhances the SF related Raman scattering~\cite{Zucker}. As depicted in Fig.~\ref{Bdep}, the e-SF intensity measured in Faraday geometry increases exponentially with the magnetic field by three orders of magnitude, while the one in Voigt geometry is only slightly enhanced with the field. The increasing magnetic field in transverse geometry, as expected, weakly affects the dimension of the exciton wave function and thus the electron spin-flip scattering efficiency.
\item[(ii)] The polarization ratio in Faraday geometry, given by Eq.~(\ref{intensratios}), amounts to $1 : 0.31 : 0.39 : 0.15$ for ($+-$):($++$):($--$):($-+$). The ratios are not sensitive to illumination, i.e., to the concentration of resident electrons.
\item[(iii)] The e-SF intensity maximum is spectrally shifted by about $\Delta E_\text{e}$ from the exciton PLE maximum, compare Fig.~\ref{fig5}(a). It is expected for a SFRS process where the outgoing resonance energy coincides with the PLE maximum of the exciton.
\item[(iv)] The e-SFRS is sensitive to illumination, shows a nonmonotonic behavior with the maximum at $P_\text{a}=0.06$~W/cm$^2$ followed by a strong decrease at higher intensities until its disappearance at $P_\text{a}=3$~W/cm$^2$, see Fig.~\ref{fig9}(b). Note that the decrease takes place at power densities where the heating of resident electrons, which would lead to a reduction in the PL polarization degree, is negligible.
\item[(v)] It shows a strong temperature dependence (Fig.~\ref{fig8}) with a deactivation energy of $0.8$~meV for the excitation at the X-maximum and $0.3$~meV at the high-energy side.
\end{itemize}

By taking into account these properties of the electron-SF at the exciton resonance, we can consider two main Raman mechanisms with high scattering probabilities. Both mechanisms are illustrated in Fig.~\ref{fig15}(b) and noted in Table~\ref{taboverview}.
\begin{itemize}
\item[(1)] \textit{Exciton only}. This mechanism does not involve resident carriers, only the electron and hole in the exciton. After $\sigma^+$ excitation, the $| +1 \rangle$ exciton state is populated. The electron spin-flip accompanied with the emission of an acoustic phonon scatters the bright exciton to the $|+2 \rangle$ state thus resulting in a Raman shift of $\Delta E_\text{e} + \delta$. The exciton exchange energy $\delta$ will not contribute if the hole spin-relaxation time $\tau_s^\text{h}$ is very short, i.e., $1/\tau_s^\text{h} \gg \delta, \omega_{\text{\tiny L}}$, so that the criterion suggested by Dyakonov et al. is fulfilled~\cite{dyakonov1997}. The fast hole relaxation mixes the hole states $|-3/2\rangle$ and $|+3/2 \rangle$, therefore, the exciton state $|+2 \rangle$ can recombine with emission of $\sigma^-$ polarized light. The criterion $1/\tau_s^\text{h} \gg \delta, \omega_{\text{\tiny L}}$ is easy to fulfill in the Voigt geometry when the Larmor frequency of the heavy-hole is low due to the small transverse component of its $g$ factor. To satisfy this criterion for the Faraday geometry and high magnetic fields ($B \approx 10$~T) the spin relaxation time should be below $1$~ps. However, this is significantly shorter than the hole spin-dephasing time of about $40$~ps measured for the sample studied~\cite{Zhukov}.

This mechanism is also in agreement with the strong difference between the e-SF intensities in Faraday and Voigt geometries which can be explained by the role of magnetic localization being stronger in the Faraday geometry. If we account for the hh-lh mixing in the studied QW with $\beta = 0.3$, the expected ratios of the different polarization configurations $1 : 0.3 : 0.3 : 0.15$ for ($+-$):($++$):($--$):($-+$) are close to the experimentally determined ratios.

The SFRS sensitivity to the illumination can be attributed to the exciton lifetime, which is limited by the exciton capture into trion. Here, we can expect a nonmonotonic dependence of the e-SF intensity on the major type of resident carrier changing from the p-type to the n-type regime, with a maximum for an empty QW. It is similar to the dependence in Fig.~\ref{fig9}(b) shown by triangles.

The temperature dependence in this case can be explained by the exciton localization. It avoids to care about the conservation of the light wave vector, which is necessary for free excitons.

\item[(2)] \textit{Only exciton + anisotropic exchange in exciton}. This mechanism consists of the e-SF via acoustic phonon scattering and anisotropic exchange between the electron and hole in the exciton. The exchange interaction, which leaves the electron spin invariant but changes the hole spin, is efficient in Faraday geometry due to similar electron and hole $g$ factors (see Subsec.~\ref{exch_int}) as well as spatially shifted hole and electron localization centers in the QW plane. This anisotropic flip-stop-like exchange interaction promotes the hole into a virtual state whose energy corresponds to the energy of the initial hole state. The virtual state is not an eigenstate of the exciton, but serves as an intermediate state in the coherent Raman scattering process. Accordingly, only the energy of the electron Zeeman splitting contributes to the Raman shift in agreement with the experimental results. The admixture of light-hole to heavy-hole states explains all polarization properties of the observed e-SF line. In comparison to model (1), the probability of the whole scattering mechanism is reduced, since the additional anisotropic e-h exchange makes it to a fourth-order Raman process.
\end{itemize}

Both processes (1) and (2) are considered as valuable candidates for explaining the e-SF at the exciton resonance. The observed strong decrease in the e-SF intensity (contrary to expectation) with increasing angle between the magnetic field and QW growth direction rather supports model (2). Indeed, both models predict an enhancement of the e-SF efficiency with the tilting angle ($I \sim \sin^2 \theta$) due to mixing of the spin-up and spin-down states by the transverse component of the magnetic field. However, only model (2) includes the competition between the above mentioned effect and the angular dependent strength of the anisotropic exchange interaction. The efficiency of the anisotropic exchange in the exciton decreases with growing angle owing to the progressively increasing mismatch between the electron and hole $g$ factors, see Eq.~(\ref{intvsangl}). On the whole, both e-SFRS models are supposed to take place for resonant excitation of the exciton. In tilted geometries, for both models the light polarization characteristics are modified by the electron-state mixing. An e-SFRS process of second order is not allowed by the mixed electron states in the case of a two carrier complex (exciton).

Note, the efficient mechanisms (1) and (2) can also serve as strong competitors for the flip of the exciton spin, which could be visible in titled geometries but was not detected. The reason for that can be its probability being lower than that of the previously described mechanisms which conceal the exciton-SFRS.

Let us discuss other mechanisms leading to the observation of e-SFRS for the resonance excitation of the exciton.

\begin{itemize}
\item[(3)] \textit{Exciton + resident electron, third-order process}. In contrast to model (1), the electron spin-flip is provided by isotropic exchange with a resident electron. In accordance with case (1), this mechanism assumes fast hole spin relaxation giving rise to a third-order Raman process whose probability is highest in cross-polarization. Although it is similar to the first mechanism, it can be distinguished through specific polarization properties when the resident electron is polarized by the external magnetic field. Such resident electron polarization in a magnetic field changes the Stokes/anti-Stokes ratio of the e-SF line intensities measured in co-polarized configurations~\cite{Akimov11}. This ratio is then not only defined by a temperature factor, but also by the electron spin polarization. For instance, the e-SF line for the ($\sigma^+,\sigma^+$) configuration might be visible only in the Stokes region, and in anti-Stokes for ($\sigma^-, \sigma^-$). However, such polarization changes were not found for the e-SF line, while the applied magnetic fields were strong enough to polarize resident electrons. Therefore, we conclude that this mechanism is not relevant.

\item[(4)] \textit{Analog of exciton + resident hole}. Another third-order mechanism is based on anisotropic exchange interaction between the photogenerated and a resident electron, by analogy with Fig.~\ref{fig15}(a). This mechanism is valid for the low-symmetry complex of a resident electron bound to the photo-exciton, where both electron wave functions are displaced. Since only the resident electron spin is changed, while the one of the photo-electron stays invariant, the process is allowed in co-polarized configurations and cannot explain the experiment for the e-SFRS.
\end{itemize}

The models (3) and (4) based on third-order processes, which have rather high scattering probabilities, predict properties of the e-SF in the exciton which are contradictory to the experimental results. Accordingly, they can be discarded, and only the models (1) and (2) are in reasonable agreement with the experiment.

\subsection{Effect of above-barrier illumination on e-SF at exciton resonance}
The enhancement of the e-SF intensity by the above-barrier illumination can be explained by an increase in the exciton lifetime. As shown by Koudinov et al.~\cite{Koudinovb}, the SFRS intensity depends on the third power of the exciton lifetime $\tau_{_\text{X}}$ or, respectively, the line width of the exciton state. Due to the background concentration of holes the excitons can become trapped in positive trion states. This trion formation is the dominant nonradiative decay mechanism for excitons. Its formation time $\tau_{_\text{T}}$ is about a few ps depending on the resident carrier concentration, while the exciton lifetime approaches 30~ps. Hence, the line width of the exciton will be broadened and, in turn, the SFRS intensity decreases. The illumination leads to a reduction in the resident hole concentration thus making the formation of positive trions less probable. Correspondingly, the lifetime of the exciton states is increased resulting in a rise in the overall electron-SFRS intensity. This rise requires that the scattering time $\tau_{\text{s}}$ of the acoustic phonon and spin-spin exchange exceeds the trion formation time, and the entire e-SFRS must occur within the exciton lifetime: $\tau_{_\text{T}} < \tau_{\text{s}} < \tau_{_\text{X}}$. At high power densities of the illumination the SFRS intensity will be reduced again, as shown in Fig.~\ref{fig9}(b). Note that the trion formation time and/or exciton lifetime only influence the SFRS intensity, while the width of the SFRS line is affected by $\tau_{\text{s}}$.

\section{Conclusion} \label{concl}
The spin-flip Raman scattering of electrons and heavy-holes has been studied for resonant excitation of neutral and charged excitons in a CdTe/Cd$_{0.63}$Mg$_{0.37}$Te quantum well. For neutral excitons SFRS models taking into account high-symmetry potential predict the flip of the exciton spin via acoustic phonon interaction in every geometry. A single spin-flip of either the electron or heavy-hole is forbidden by selection rules in the electric-dipole approximation. However, in the Raman spectra measured only the single spin-flips are observed. For the heavy-hole-SF, the anisotropic flip-stop-like exchange interaction between the heavy-hole in the exciton and a neighbored heavy-hole comes into consideration as scattering mechanism. Due to spatially mismatched wave functions of both holes in the QW plane, reflecting a reduced QW potential symmetry, the anisotropic exchange is possible. Hereby, only the resident heavy-hole spin state is changed giving rise to a Raman shift corresponding to $g_{\text{hh}}$.

The electron spin-flip scattering in the neutral exciton is provided by two mechanisms. They shall describe the three main features of the e-SF derived from experiment: its strong intensity in crossed polarization configuration, intensity enhancement with increasing magnetic field in Faraday geometry, and a decrease in intensity by going from Faraday to Voigt geometry. In both mechanisms, the electron spin is flipped via acoustic phonon interaction. In order to explain the crossed polarization, model (1) assumes a fast hole spin relaxation, and model (2) supposes an anisotropic electron-hole exchange interaction. The first model is of lower scattering order, hence, its probability is higher than that of the second model. But, model (1) requires a fast hole spin relaxation, and cannot describe the e-SF intensity reduction with increasing tilting angle, in contrast to model (2). Here, the anisotropic exchange interaction looses its strength due to the increasing difference between the electron and heavy-hole Larmor frequencies with growing angle. The intensity enhancement with magnetic field strength is attributed to a stronger magnetic localization giving rise to larger spin dephasing and life times of the exciton. The circular polarization characteristics of the SFRS lines are fully described in consideration of light-heavy-hole mixing, inherently induced by local QW deformation and asymmetric interface potential.

For positively and negatively charged excitons, the simultaneous spin-flips of electron and heavy-hole are expected from the SFRS models in Faraday geometry. Nevertheless, only an e-SF for the T$^-$ trion is observed. It is based on anisotropic electron-heavy-hole exchange interaction, which opens the outgoing scattering channel for circularly polarized recombination. In tilted geometry, the models allow electron-SF in T$^-$ trion and heavy-hole-SF in T$^+$ trion due to field-induced electron-state mixing. These efficient Raman processes of second order are confirmed by experiment.

In the two-color SFRS measurements with resonant excitation of exciton complexes and above-barrier illumination, we have shown that the intensity of the electron-SFRS can be considerably enhanced by moderate illumination power densities. Too high power densities lead to reduction in the exciton lifetime and delocalization of resident electrons. From the temperature dependences of the SFRS lines, the deactivation energies of the corresponding spin-flip Raman processes have been evaluated. They are strongly connected to the localization energies of the respective exciton complexes. Besides the lifetime of the exciton complex under study, also the localization degree of the carrier involved in the scattering process control the SFRS efficiency.

We have demonstrated that the spin-flip Raman scattering gives valuable information not only about the electron and heavy-hole $g$ factor tensors, but particularly about the spin-flip scattering mechanisms being present in neutral and charged exciton complexes. Besides the dependence of the spin-flip scattering mechanisms on the strength of the magnetic field, lattice temperature and laser power, especially the geometry of experiment, the polarization properties of the incident and scattered light as well as the excitation energy are crucial parameters to observe and characterize SFRS processes.

\section*{Acknowledgments}
We greatly acknowledge E.~L.~Ivchenko for fruitful discussions, and D.~Braukmann as well as D.~Kudlacik for the help with experiment. We also thank R.~Sch\"afer from Spectroscopy \& Imaging GmbH for assistance with the TriVista triple-spectrometer. This work was supported by the Deutsche Forschungsgemeinschaft via SPP 1285, the EU Seventh Framework Programme (Grant No. 237252, Spin-optronics), the Russian Foundation for Basic Research, and the Mercator Research Center Ruhr (MERCUR) of Stiftung Mercator. The research stay of VFS in Dortmund was supported by the Deutsche Forschungsgemeinschaft (Grant No. YA65/10-1).


\begin{thebibliography}{}
\bibitem{notegfactor} The $g$ factor signs cannot be directly evaluated from the SFRS experiments. However, it is known that in the studied CdTe/(Cd,Mg)Te QW both electron and heavy-hole $g^{\parallel}$ factors are negative \cite{Bartsch}.
\end{thebibliography}
\end{document}